# Engineering phase-frustration induced flat bands in an aza-triangulene covalent Kagome lattice

Yuyi Yan[1]†, Fujia Liu[1]†, Weichen Tang[2,3]†, Han Xuan Wong[1], Boyu Qie[1,4,5]*, Steven G. Louie[2,3]*, Felix R. Fischer[1,3,4,5]*

[1]Department of Chemistry, University of California, Berkeley, CA 94720, USA. [2]Department of Physics, University of California, Berkeley, CA 94720, USA. [3]Materials Sciences Division, Lawrence Berkeley National Laboratory, Berkeley, CA 94720, USA. [4]Kavli Energy NanoSciences Institute at the University of California Berkeley and the Lawrence Berkeley National Laboratory, Berkeley, California 94720, USA. [5]Bakar Institute of Digital Materials for the Planet, Division of Computing, Data Science, and Society, University of California Berkeley; Berkeley, CA 94720, USA.

† These authors contributed equally to this work.

* Corresponding authors.

$\pi$-Conjugated covalent organic frameworks (COFs) provide a versatile platform for the realization of designer quantum nanomaterials. Strong electron–electron correlation within these artificial lattices can give rise to exotic phases of matter. Their experimental realization however requires precise control over orbital symmetry, charge localization, and band dispersion all arising from the effective hybridization between molecular linkers and nodes. Here, we present a modular strategy for constructing diatomic Kagome lattices from aza-[3]triangulene (A[3]T) nodes, in which a $D_{3h}$ symmetric ground state is stabilized through resonance contributions from a cumulenenic linker. First-principles density-functional theory and scanning tunnelling spectroscopy reveal that the hybridization of a sixfold degenerate set of edge-localized Wannier functions in the unit cell gives rise to orbital-phase frustration-induced non-trivial flat bands. These results establish a general design principle for engineering orbital interactions in organic lattices and open a pathway toward programmable COF-based quantum materials with correlated electronic ground states.

Flat-band systems that confine electronic states to narrow energy bands near the Fermi level ($E_F$) have garnered interest in the exploration of low-dimensional many-body quantum phenomena[1-6]. Unlike trivial flat bands (FBs) associated with localized atomic or molecular orbitals, phase frustration-induced FBs[7] arise naturally from a quenching of the electron dispersion through destructive wave function interference that dramatically reduces the hopping amplitude in select classes of frustrated lattices, such as Kagome, Lieb, dice, and multi-orbital honeycomb lattices[8-10]. Increased correlation (that is, potential over kinetic energy) of carriers in these non-dispersive band systems can lead to the emergence of unconventional quantum phenomena that combine many-body interactions with topological effects, giving rise to exotic behaviours in superconductivity[11,12], magnetism[13], excitonic insulators[14], Wigner crystals[15], and manifestations of the fractional quantum Hall effect[16,17]. The complex challenges associated with engineering orbital interactions that give rise to phase frustration-induced FBs in covalent molecular lattices, along with the requirement to preclude spontaneous lattice distortions that break the symmetry of open-shell molecular building blocks (e.g. Jahn-Teller like distortions[18]), represent a veritable challenge to the fabrication and exploration of carbon-based quantum materials.

Here we report the modular design and on-surface synthesis of a π-conjugated aza-[3]triangulene COF (A[3]TCOF) assembled from N-atom doped [3]triangulene cores (the nodes on the COF lattice) linked by ethynylene groups (–C≡C–) (the linkers on the COF lattice) fused normal to their zigzag edges (Figure 1a). The electronic structure of such an edge-sharing 2D-A[3]T crystal can be modelled as a diatomic Kagome lattice[19-21]. In this picture, each A[3]T core is represented as a superatom (triangle) hosting a set of three symmetry related single-occupied frontier states (blue circles along the triangle edges in Figure 1b). Hybridization between frontier states on adjacent A[3]T cores gives rise to a pair of valence band (VB) and conduction band (CB) complexes (each comprised of three bands) that bracket the $E_F$. Differential conductance spectroscopy of extended A[3]TCOF lattices revealed the distinctive signatures of Kagome FBs near the $E_F$, consistent with density functional theory (DFT) simulations and Wannier function (WF) analysis. This highly modular and general approach for designing and accessing Kagome flat band materials within a COF using the tools of on-surface synthesis paves the way for the realization and exploration of strongly correlated and higher order topological phases in exotic carbon-based quantum materials.

**Design of phase-frustrated Kagome flat bands in A[3]TCOFs**

Our strategy for designing phase-frustrated FBs within the framework of A[3]TCOFs is based on a simple nearest-neighbour (NN) tight-binding (TB) model. Whereas the ground state spin-multiplicity of all-carbon [$n$]triangulenes can be predicted by Ovchinnikov's rule[22,23] and Lieb's theorem[24] for bipartite lattices ($S$ = ½ ($N_A - N_B$), where $N_A$ and $N_B$ represent the number of atoms residing on the intrinsic A and B graphene sublattices, respectively), the substitutional doping in A[$n$]Ts requires a more detailed analysis. Density functional theory (DFT) calculations (M06-2X/def2-TZVP)[25-29] predict that the electronic ground state of the parent A[3]T core is best represented as an open-shell doublet arising from a Jahn-Teller like distortion of the molecular structure from $D_{3h}$ to $C_{2v}$ (Figure 1c, Extended Data Figure ED1)[30,31]. Strong hybridization between edge-fused A[3]T cores mediated by sizeable contributions from a cumulenic resonance structure[32-34] (Figure 1d) stabilizes the $D_{3h}$ symmetric A[3]T ground state (over the $C_{2v}$ symmetric A[3]T) and restores the requisite set of three single-occupied frontier orbitals per lattice node. Thus, we can apply a simple electron hopping model for a periodic 2D lattice of zigzag edge linked A[3]Ts (Figure 1b) to the construction of a low-energy effective TB Hamiltonian for spinless electrons

$$\hat{\mathcal{H}} = -t_1 \sum_{\langle ij \rangle_{intra}} c_i^\dagger c_j - t_2 \sum_{\langle ij \rangle_{inter}} c_i^\dagger c_j + \varepsilon_A \sum_i c_i^\dagger c_i$$

where $t_1$, and $t_2$ are the intra- and inter-A[3]T electron hopping parameters, $\varepsilon_A$ is the onsite energy of localized zigzag edge states, and $c_i^\dagger$ ($c_j^\dagger$) and $c_i$ ($c_j$) are the creation and annihilation operators for electrons at site $i$ and $j$, respectively.

A plot of the resulting band diagram (Figure 1e) shows the expected conduction band (CB) and valence band (VB) complexes of a so-called Type-II diatomic Kagome lattice (pairs of dispersive Dirac bands bounded by FBs at lower energy for both the CB and VB complexes)[19-21]. Interference patterns arising from destructive wave function overlap — nodal lines in Figure 1f bisect the A[3]T cores localizing the corresponding wave functions along the inner circumference of the pores of the lattice — give rise to two phase frustration-induced Kagome FBs, one each at the bottom of the CB and VB complex, respectively.

**Synthesis of A[3]TCOFs**

Guided by this idea, we designed a molecular precursor for the on-surface synthesis of A[3]TCOFs. The aza-[3]triangulene **1** (Figure 2a) could be derived in a single step from $4H$-benzo[9,1]quinolizino[3,4,5,6,7-$defg$]acridine-4,8,12-trione through a Ramirez reaction with carbon tetrachloride[35,36]. The reaction yielded a mixture of the desired trisubstituted aza-[3]triangulene **1** along with the lower molecular weight disubstituted aza-[3]triangulene **2** that retained one of the original carbonyl groups (Extended Data Figure ED2). Molecular precursors **1** and **2** were each sublimed from a molecular beam evaporator (MBE) in ultrahigh vacuum (UHV) onto clean Au(111) substrates held at 297 K. Constant-current scanning tunnelling microscopy (STM) images ($T = 4$ K) of densely packed self-assembled islands of **1** and **2** on Au(111) are shown in the Extended Data Figure ED3. Annealing of molecule decorated surfaces to $T = 523$ K at a rate of 3 K min$^{-1}$ induced the homolytic cleavage of thermally labile C–Cl bonds lining the edges of **1** or **2** followed by intermolecular radical recombination of the surface-stabilized vinylidene-carbene intermediates yielding the corresponding A[3]T networks fused by cumulene/ethynylene linkers[32,37,38]. While the trisubstituted precursor **1** gave rise to the Kagome lattice of a A[3]TCOF the disubstituted aza-[3]triangulene **2** assembled into extended linear chains and cyclic hexamers.

Figure 2b shows a representative topographic STM image of a locally ordered A[3]TCOF island (~160 nm$^2$) with an apparent uniform height of $0.20 \pm 0.03$ nm. BRSTM and non-contact AFM (nc-AFM) imaging with CO-functionalized tips confirmed the formation of the expected A[3]T framework devoid of skeletal rearrangements. Closer inspection of a greater than 4 × 4 cell (Figure 2c) revealed the characteristic *p6mm* symmetry of A[3]TCOFs. Each triangulene core is fused along the zigzag edges via cumulene/ethynylene linkers to three nearest neighbours in the hexagonal lattice (lattice constants $|a_1| = |a_2| = 1.76 \pm 0.05$ nm; $\alpha = 60°$) (Figure 2d, Extended Data Figure ED4).

Annealing of surfaces decorated with the difunctional A[3]T precursor **2** yielded a mixture of linear A[3]T chains (Figure 2e) and cyclic (A[3]T)$_6$ (Figure 2f). Linear A[3]T chains containing more than six repeat units tended to be disordered and adopted a random sequence of s-*cis* and s-*trans* conformations that place the carbonyl groups along both edges of the chain (Figure 2g, Extended Data Figure ED4). The end groups of these linear A[3]T chains commonly featured both a C=O along with a vinylidene group resulting from hydrogen transfer to the highly reactive carbene intermediate (Figure 2h and i). Shorter oligomers comprised of six or fewer monomer units adopted a minimally strained all s-*cis* conformation that favours the fusion of chain ends into six-membered rings, *i.e.*, (A[3]T)$_6$. The outer perimeter of (A[3]T)$_6$ was lined by carbonyl and $C_{aryl}$–H groups that drive the self-assembly into islands of densely packed hexagonal lattices (lattice constants $|b_1| = |b_2| = 2.85 \pm 0.05$ nm; $\beta = 60°$) (Figure 2j and k, Extended Data Figure ED5).

**Electronic structure of A[3]T lattices**

The electronic structure of 2D A[3]TCOFs was characterized by differential conductance spectroscopy (for A[3]T chains and (A[3]T)$_6$ see Supplementary Information Figures S1–5). d$I$/d$V$ point spectra recorded at the positions marked in the BRSTM inset in Figure 3a, part of a locally ordered island (10 nm × 13 nm) containing > 50 unit cells, showed six prominent features. A broad peak centred at $V_s = +1.30 \pm 0.05$ V (*feature* 1), a shoulder at $V_s = +0.80 \pm 0.05$ V (*feature* 2) and a sharp peak at $V_s = +0.55 \pm 0.05$ V (*feature* 3) dominated d$I$/d$V$ point spectra recorded above the position of the cumulene/ethynylene linkers (the intensity was only slightly reduced above the A[3]T cores). At negative bias a set of three distinctive signals could be observed: a small peak at $V_s = –0.05 \pm 0.03$ V (*feature* 4), a shoulder at $V_s = –0.15 \pm 0.05$ V (*feature* 5), and a sharp peak at $V_s = –0.30 \pm 0.05$ V (*feature* 6) that rapidly decayed into the signal background.

d$I$/d$V$ maps recorded in incremental steps across a large bias window (+1.50 V > $V_s$ > –1.50 V) resolved the distinctive signatures of electronic states arising from the A[3]TCOF lattice (Extended Data Figure ED6, a second dataset is included in Supplementary Information Figure S6). At positive bias above $V_s > +1.2$ V across the width of *feature* 1 d$I$/d$V$ maps showed a distinct nodal pattern, a large signal intensity on the triangulene cores (Figure 3b), that gradually transitioned into a striped electron density (+1.2 V > $V_s$ > +0.9 V) that extended across both cores and linkers (Figure 3c). At $V_s \sim +0.8$ V, the position of a shoulder (*feature* 2) in the d$I$/d$V$ point spectrum, a prominent bowtie nodal pattern emerges that coincides with the position of the cumulene/ethynylene linkers. This nodal pattern continued to dominate d$I$/d$V$ maps spanning across near the full width of *feature* 3 (Figure 3d), the peak centred at $V_s = +0.55$ V, before sharply decaying into a diffuse signal background. Across the bias window from +0.45 V > $V_s$ > +0.05 V, d$I$/d$V$ maps remained featureless, indicative of an energy gap ($E_{g,Kagome,exp} \sim 0.40$ V) adjacent to the $E_F$ (Extended Data Figure ED6i–n). Below $V_s \sim +0.05$ V d$I$/d$V$ maps featured a distinctive dark trefoil nodal pattern cantered on each A[3]T core (Figure 4e) that persisted uninterrupted across $E_F$ and merged into *feature* 4 ($V_s = -0.05$ V). Shifting to more negative bias, the nodal pattern in d$I$/d$V$ maps underwent sudden dramatic transformations. Over a span less than 50 mV, the trefoil shapes evolved into a striated web-like pattern that spanned the width of both the shoulder (*feature* 5) and most of the prominent peak (*feature* 6) at $V_s = -0.15$ V and $V_s = -0.30$ V, respectively (Figure 4f). It is only at the very tail-end of *feature* 6, over a narrow bias window (–0.30 > $V_s$ > –0.40 V), that the nodal pattern in d$I$/d$V$ maps changed one last time to reveal a geometric petal-like arrangement of bright lobes evenly distributed across both A[3]T cores and linkers (Figure 3g).

Experimental d$I$/d$V$ spectra and maps were reproduced by theory based on *ab initio* DFT calculations using the Perdew-Burke-Ernzerhof (PBE) functional[39]. The computed density of states (DOS) (Figure 3n) and band structure (Figure 3o) of a freestanding A[3]TCOF showed the characteristic features of CB and VB complexes expected for a diatomic Type-II Kagome lattice (for DFT-HSE06 results see Supplementary Information Figure S3b). Theory predicts that freestanding A[3]TCOFs are semiconductors with an indirect band gap ($\Gamma \rightarrow K$) of $E_{g,Kagome,DFT} = 0.26$ eV. The DOS is dominated by a pair of sharp peaks directly associated with the narrow band dispersion of Kagome valence (VFB) and conduction flat bands (CFB). A secondary set of broader features arises from contributions of Van Hove singularities (VHSs) at $\boldsymbol{k} = M$, critical points in the dispersive bands of VB and CB complexes (VHS1–4 in Figure 3o). Projections of the local density of states (LDOS) maps at energies corresponding to the VHS4 (Figure 3h), VHS3 (Figure 3i), CFB (Figure 3j), VHS2 (Figure 3k), VHS1 (Figure 3l), and VFB (Figure 3m) simulated using a model for $p$-wave or $s$-wave STM tips[40] were in good agreement with d$I$/d$V$ maps attributed to spectral *features* 1–6, respectively. The apparent discrepancy between experiment (partially filled bands cross $E_F$) and theory (indirect semiconducting gap) in the placement of the Fermi level suggests that the interactions with the underlying Au(111) surface induce a p-type doping in the A[3]TCOF, shifting the $E_F$ to lower energy into the dispersive Kagome VB complex.

Besides the characteristic changes to nodal patterns in d$I$/d$V$ maps recorded across a bias range of –0.45 V < $V_s$ < +1.50 V, large area scans of locally ordered A[3]TCOF lattice show long-wavelength (larger than one unit cell) modulations to the real-space electronic structure (see 5 × 8 lattice in Figure 3p). These quasiparticle interference (QPI) patterns result from electron scattering near edges or point defects (measured interference patterns emerge only from the electronic structure and are absent in nc-AFM images, Extended Data Figure ED4,d). Fourier transform (FT) of the corresponding differential conductance maps gave access to the energy resolved band structure in $\boldsymbol{k}$-space[41,42]. Figure 3q shows the FT of a d$I$/d$V$ map recorded at a bias of $V_s$ –0.30 V (see Extended Data Figure ED7 for full dataset; a second dataset on a 5 × 6 lattice is included in Supplementary Information Figure S7). The FT maps are dominated by the sixfold symmetric Bragg peaks ($b_i$, 1 ⩽ $i$ ⩽ 6) of the reciprocal lattice. The first Brillouin zone (BZ) (dashed hexagon) features two additional peaks at small $\boldsymbol{k}$-values along the $M$–$\Gamma$–$M$ direction that evolve with the applied sample bias. The reduced symmetry of FT maps can be attributed to a preferred axis of propagation for QPIs within finite (< 50 unit cells) islands of A[3]TCOF (experimentally accessible A[3]TCOF lattices are too small to capture a real-space averaged signal propagating along all lattice vectors). We initially evaluated the total normalized intensity ($\sum_{BZ} \frac{I}{I_0}$) and the median normalized intensity ($\mu_{1/2}$) in the first BZ, proportional to the global DOS, as a function of energy for empty (Figure 3r) and filled states (Figure 3s). The most prominent peaks coincided with the position of the CFB (*feature* 3, $V_s$ ~ +0.60

V) and the VFB (*feature 6*, $V_s \sim -0.30$ V) derived from d$I$/d$V$ point spectra. Lesser peaks could be mapped onto the position of the VHS1 (*feature 5*, $V_s \sim -150$ mV), VHS2 (*feature 4*, $V_s \sim -25$ mV), VHS3 (*feature 2*, $V_s \sim +750$ mV), and VHS4 (*feature 1*, $V_s \sim +1250$ mV), respectively.

The energy and momentum resolved electronic structure for empty and filled states extracted from FT maps is shown in Figures 3t and 3u, respectively. The averaged FT signal amplitude as a function of wave vector ***k*** plotted along the $\Gamma$–$K$–$M$–$\Gamma$ path (very coarse owing to the small size and irregular shape of samples) shows two distinct features: At positive bias a sharp signal onset close to the $\Gamma$-point ($V_s \sim +0.55$ V) coincided with the position of *feature 3* the CFB derived from d$I$/d$V$ point spectroscopy (Figures 3t). While theory predicts a uniform distribution of the DOS at energies corresponding to the FB, the signal intensity gradually fades along the path from $\Gamma$–$K$ before reemerging closer to the $M$-point. Rather than suggesting an electronic effect, the observed gradient follows the reduced sensitivity of constant-height STS when probing states associated with large contributions from parallel momentum ***k***$_\parallel$ (the tunnelling matrix element favours electronic states with small ***k***$_\parallel$) and is superimposed on the signal associated with the CFB[43,44]. As the bias increases the signal gradually spreads to larger values of ***k*** reaching the $M$-point at $V_s \sim +0.80$ V (blue dashed oval) consistent with the expected position of the VHS3 (*feature 2*). A similar picture could be observed for the occupied states. Approaching from negative bias the FT signal at small ***k*** (close to $\Gamma$) emerges above the background at $V_s \sim -0.30$ V (Figures 3u). With increasing bias, the signal broadened towards the $M$-point ($-225$ mV $< V_s < -125$ V, blue dashed oval), the position of VHS1. Above $V_s = -0.10$ V the signal receded back toward the $\Gamma$-point before vanishing into the background at positive bias. Despite significant experimental constraints imposed by the finite size of A[3]TCOF samples coupled with an unusually large primitive unit cell ($A_p = 2.68$ nm$^2$) the experimental observation of characteristic features associated with the VFB, CFB, and the VHSs in momentum and energy resolved STS supports our theoretical model that describes the emergence of phase frustration-induced non-trivial Kagome FBs in edge-linked A[3]TCOFs.

## Bonding in edge-linked A[3]T lattices

While isolated A[3]Ts have been shown to host unpaired spins, experiment and theory showed that the strongly hybridized edge linked A[3]TCOFs are closed-shell (diamagnetic) structures (for a model of the cumulene linked A[3]T dimer see Extended Data Figure ED8). A transformation of the periodic Bloch states into the corresponding degenerate maximally localized Wannier functions (MLWFs) yielded a sixfold degenerate set of single-occupied frontier states per unit cell (Figure 4a). Every zigzag edge fused through cumulene/ethynylene linkers hosts one MLWF, mirroring the $D_{3h}$ symmetry of the A[3]T core. Fitting a truncated NNN TB Hamiltonian in the Wannier basis (Figure 4b) to the DFT calculated band structure (red lines in Figure 3o) gave hopping parameters $t_1 = 0.34$ eV, $t_2 = 0.63$ eV, $t_3 = 0.04$ eV, with an onsite energy of $\varepsilon_A = 0.08$ eV. Translated into a chemical valence bond picture the strong hybridization between MLWFs on adjacent zigzag edges ($t_2 > t_1 \gg t_3$) is dominated by contributions from a cumulenic resonance structure. DFT geometry optimization of A[3]TCOF lattices predicts a bond length modulation along the linker ($C_1$–$C_2 = 1.371$ Å, $C_2$–$C_3 = 1.246$ Å, $C_3$–$C_4 = 1.371$ Å) consistent with the structure of symmetrically substituted [3]cumulenes[45,46].

An extension of this diatomic Kagome model highlights the importance of setting the relative ratios between the hopping integrals ($t_1$, $t_2$, $t_3$) to access phase-frustration induced Kagome FBs in COFs by chemical design of the linker and the node. Figure 4c shows a NNN TB phase diagram of semiconducting (Type–I–IV) and mixed metallic Kagome band structures as a function of $t_2/t_1$ and $t_3/t_1$ (colour gradient indicates magnitude of the band gap in $|E_g$ (eV)$/t_1|$; see Extended Data Figure ED9 for representative examples)[19]. The A[3]TCOF featuring a cumulene/ethynylene linker falls within the Type-II phase sharing the characteristics of two dispersive Dirac bands bounded by a flat band at lower energy for both the VB and CB complex. Ab initio DFT-PBE calculations predict that extension of the linker in A[3]TCOFs to longer polyynes (*e.g.*, buta-1,3-diyne, hexa-1,3,5-triyne, or octa-1,3,5,7-tetrayne) would shift the structures toward smaller $t_2/t_1$ (inter-A[3]T hopping $t_2$ decreases with distance while $t_1$ remains largely unchanged) along an imaginary line that crosses the Type-I/Type-II phase boundary (Figure 4c, Extended Data Figure

ED10). The reduction in $t_2$ with increasing linker length leads to a decrease of the semiconducting band gap approaching a metallic state (the Type-I/Type-II phase boundary) where the dispersive bands of either the VB or CB complexes cross $E_F$. A phase transition to the Kagome Type-I band structure can be observed if the linker in A[3]TCOFs is replaced by a 1,4-diethynylbenzene group.

Our results not only demonstrate the construction of extended carbon-based nanomaterials based on the effective stabilization of degenerate MLWF in $D_{3h}$ symmetric A[3]T building blocks but provide a blueprint for the design and chemical engineering of phase-frustration induced FBs in covalent organic Kagome lattices. This highly modular bottom-up approach paves the way for the development of nanoscale electrical devices and the exploration of strongly correlated phenomena in organic topological flat band materials.

## References

1    Kang, M. G. *et al.* Topological flat bands in frustrated kagome lattice CoSn. *Nature Communications* **11**, 4004 (2020). https://doi.org/10.1038/s41467-020-17465-1

2    Calugaru, D. *et al.* General construction and topological classification of crystalline flat bands. *Nat Phys* **18**, 185–189 (2022). https://doi.org/10.1038/s41567-021-01445-3

3    Regnault, N. *et al.* Catalogue of flat-band stoichiometric materials. *Nature* **603**, 824–828 (2022). https://doi.org/10.1038/s41586-022-04519-1

4    Guo, Q. B. *et al.* Ultrathin quantum light source with van der Waals NbOCl$_2$ crystal. *Nature* **613**, 53–59 (2023). https://doi.org/10.1038/s41586-022-05393-7

5    Ye, L. D. *et al.* Hopping frustration-induced flat band and strange metallicity in a kagome metal. *Nat Phys* **20**, 610–614 (2024). https://doi.org/10.1038/s41567-023-02360-5

6    Checkelsky, J. G., Bernevig, B. A., Coleman, P., Si, Q. M. & Paschen, S. Flat bands, strange metals and the Kondo effect. *Nat Rev Mater* **9**, 509–526 (2024). https://doi.org/10.1038/s41578-023-00644-z

7    Bergman, D. L., Wu, C. J. & Balents, L. Band touching from real-space topology in frustrated hopping models. *Physical Review B* **78**, 125104 (2008). https://doi.org/10.1103/PhysRevB.78.125104

8    Morell, E. S., Correa, J. D., Vargas, P., Pacheco, M. & Barticevic, Z. Flat bands in slightly twisted bilayer graphene: Tight-binding calculations. *Physical Review B* **82**, 121407 (2010). https://doi.org/10.1103/PhysRevB.82.121407

9    Jovanovic, M. & Schoop, L. M. Simple Chemical Rules for Predicting Band Structures of Kagome Materials. *Journal of the American Chemical Society* **144**, 10978–10991 (2022). https://doi.org/10.1021/jacs.2c04183

10   Neves, P. M. *et al.* Crystal net catalog of model flat band materials. *Npj Comput Mater* **10**, 39 (2024). https://doi.org/10.1038/s41524-024-01220-x

11   Micnas, R., Ranninger, J. & Robaszkiewicz, S. Superconductivity in Narrow-Band Systems with Local Nonretarded Attractive Interactions. *Rev Mod Phys* **62**, 113–171 (1990).   https://doi.org/DOI 10.1103/RevModPhys.62.113

12   Aoki, H. Theoretical Possibilities for Flat Band Superconductivity. *J Supercond Nov Magn* **33**, 2341–2346 (2020). https://doi.org/10.1007/s10948-020-05474-6

13   Mielke, A. Ferromagnetic Ground-States for the Hubbard-Model on Line Graphs. *J Phys a-Math Gen* **24**, L73–L77 (1991). https://doi.org/10.1088/0305-4470/24/2/005

14   Sethi, G., Zhou, Y. N., Zhu, L. H., Yang, L. & Liu, F. Flat-Band-Enabled Triplet Excitonic Insulator in a Diatomic Kagome Lattice. *Phys. Rev. Lett.* **126**, 196403 (2021). https://doi.org/10.1103/PhysRevLett.126.196403

15   Shayegan, M. Wigner crystals in flat band 2D electron systems. *Nat Rev Phys* **4**, 212–213 (2022). https://doi.org/10.1038/s42254-022-00444-4

16   Tang, E., Mei, J. W. & Wen, X. G. High-Temperature Fractional Quantum Hall States. *Physical Review Letters* **106**, 236802 (2011). https://doi.org/10.1103/PhysRevLett.106.236802

17   Bergholtz, E. J. & Liu, Z. Topological Flat Band Models and Fractional Chern Insulators. *Int J Mod Phys B* **27**, 1330017 (2013). https://doi.org/10.1142/S021797921330017x

18   Lawrence, J. *et al.* Topological Design and Synthesis of High-Spin Aza-triangulenes without Jahn-Teller Distortions. *ACS Nano* **17**, 20237-20245 (2023). https://doi.org/10.1021/acsnano.3c05974

19  Zhou, Y. N., Sethi, G., Zhang, C., Ni, X. J. & Liu, F. Giant intrinsic circular dichroism of enantiomorphic flat Chern bands and flatband devices. *Physical Review B* **102**, 125115 (2020). https://doi.org/10.1103/PhysRevB.102.125115

20  Guo, X., Mu, H. M., Hu, T. Y., Li, Q. X. & Wang, Z. F. Bipolar semiconductor in two-dimensional covalent organic frameworks. *Physical Review B* **105**, 155415 (2022). https://doi.org/10.1103/PhysRevB.105.155415

21  Hu, T. Y., Zhang, T. F., Mu, H. M. & Wang, Z. F. Intrinsic Second-Order Topological Insulator in Two-Dimensional Covalent Organic Frameworks. *J Phys Chem Lett* **13**, 10905–10911 (2022). https://doi.org/10.1021/acs.jpclett.2c02683

22  Ovchinnikov, A. A. Multiplicity of Ground-State of Large Alternant Organic-Molecules with Conjugated Bonds. *Theor Chim Acta* **47**, 297–304 (1978). https://doi.org/Doi 10.1007/Bf00549259

23  Yazyev, O. V. Emergence of magnetism in graphene materials and nanostructures. *Rep. Prog. Phys.* **73**, 056501 (2010). https://doi.org/10.1088/0034-4885/73/5/056501

24  Lieb, E. H. Two Theorems on the Hubbard-Model. *Phys. Rev. Lett.* **62**, 1201–1204 (1989). https://doi.org/10.1103/PhysRevLett.62.1201

25  Zhao, Y. & Truhlar, D. G. The M06 suite of density functionals for main group thermochemistry, thermochemical kinetics, noncovalent interactions, excited states, and transition elements: two new functionals and systematic testing of four M06-class functionals and 12 other functionals. *Theor Chem Acc* **120**, 215–241 (2008). https://doi.org/10.1007/s00214-007-0310-x

26  Lu, T. & Chen, F. W. Multiwfn: A multifunctional wavefunction analyzer. *J Comput Chem* **33**, 580–592 (2012). https://doi.org/10.1002/jcc.22885

27  Neese, F. The ORCA program system. *Wires Comput Mol Sci* **2**, 73–78 (2012). https://doi.org/10.1002/wcms.81

28  Weigend, F. & Ahlrichs, R. Balanced basis sets of split valence, triple zeta valence and quadruple zeta valence quality for H to Rn: Design and assessment of accuracy. *Phys Chem Chem Phys* **7**, 3297–3305 (2005). https://doi.org/10.1039/b508541a

29 Lu, T. A comprehensive electron wavefunction analysis toolbox for chemists, Multiwfn. *J Chem Phys* **161**, 082503 (2024). https://doi.org/10.1063/5.0216272

30 Jahn, H. A. & Teller, E. Stability of polyatomic molecules in degenerate electronic states. I. Orbital degeneracy. *Proc R Soc Lon Ser-A* **161**, 220–235 (1937). https://doi.org/10.1098/rspa.1937.0142

31 Jahn, H. A. Stability of polyatomic molecules in degenerate electronic states II – Spin degeneracy. *Proc R Soc Lon Ser-A* **164**, 0117–0131 (1938). https://doi.org/10.1098/rspa.1938.0008

32 Cirera, B. *et al.* Tailoring topological order and π-conjugation to engineer quasi-metallic polymers. *Nature Nanotechnology* **15**, 437–443 (2020). https://doi.org/10.1038/s41565-020-0668-7

33 González-Herrero, H. *et al.* Atomic Scale Control and Visualization of Topological Quantum Phase Transition in π-Conjugated Polymers Driven by Their Length. *Adv Mater* **33**, 2104495 (2021). https://doi.org/10.1002/adma.202104495

34 Jiménez-Martín, A. *et al.* Atomically Precise Control of Topological State Hybridization in Conjugated Polymers. *ACS Nano* **18**, 29902–29912 (2024). https://doi.org/10.1021/acsnano.4c10357

35 Ramirez, F., Mckelvie, N. & Desai, N. B. New Synthesis of 1,1-Dibromoolefins Via Phosphine-Dibromomethylenes - Reaction of Triphenylphosphine with Carbon Tetrabromide. *Journal of the American Chemical Society* **84**, 1745–1747 (1962). https://doi.org/10.1021/ja00868a057

36 Giguère, J. B., Verolet, Q. & Morin, J. F. 4,10-Dibromoanthanthrone as a New Building Block for p-Type, n-Type, and Ambipolar π-Conjugated Materials. *Chemistry-a European Journal* **19**, 372–381 (2013). https://doi.org/10.1002/chem.201202878

37 Sun, Q. *et al.* On-Surface Formation of Cumulene by Dehalogenative Homocoupling of Alkenyl-Dibromides. *Angewandte Chemie-International Edition* **56**, 12165–12169 (2017). https://doi.org/10.1002/anie.201706104

38 Sánchez-Grande, A. *et al.* On-Surface Synthesis of Ethynylene-Bridged Anthracene Polymers. *Angewandte Chemie-International Edition* **58**, 6559–6563 (2019). https://doi.org/10.1002/anie.201814154

39 Perdew, J. P., Burke, K. & Ernzerhof, M. Generalized gradient approximation made simple. *Physical Review Letters* **77**, 3865–3868 (1996). https://doi.org/10.1103/PhysRevLett.77.3865

40 Gross, L. *et al.* High-Resolution Molecular Orbital Imaging Using a *p*-Wave STM Tip. *Physical Review Letters* **107**, 086101 (2011). https://doi.org/10.1103/PhysRevLett.107.086101

41 Crommie, M. F., Lutz, C. P. & Eigler, D. M. Imaging Standing Waves in a 2-Dimensional Electron-Gas. *Nature* **363**, 524–527 (1993). https://doi.org/10.1038/363524a0

42 Hasegawa, Y. & Avouris, P. Direct Observation of Standing-Wave Formation at Surface Steps Using Scanning Tunneling Spectroscopy. *Physical Review Letters* **71**, 1071–1074 (1993). https://doi.org/10.1103/PhysRevLett.71.1071

43 Tersoff, J. & Hamann, D. R. Theory and Application for the Scanning Tunneling Microscope. *Physical Review Letters* **50**, 1998-2001 (1983). https://doi.org/10.1103/PhysRevLett.50.1998

44 Tersoff, J. & Hamann, D. R. Theory of the Scanning Tunneling Microscope. *Physical Review B* **31**, 805-813 (1985). https://doi.org/10.1103/PhysRevB.31.805

45 Januszewski, J. A., Wendinger, D., Methfessel, C. D., Hampel, F. & Tykwinski, R. R. Synthesis and Structure of Tetraarylcumulenes: Characterization of Bond-Length Alternation versus Molecule Length. *Angewandte Chemie-International Edition* **52**, 1817-1821 (2013). https://doi.org/10.1002/anie.201208058

46 Gawel, P. *et al.* Synthesis of Cyano-Substituted Diaryltetracenes from Tetraaryl[3] cumulenes. *Angewandte Chemie-International Edition* **53**, 4341-4345 (2014). https://doi.org/10.1002/anie.201402299

47 Giannozzi, P. *et al.* QUANTUM ESPRESSO: a modular and open-source software project for quantum simulations of materials. *J. Phys. Condens. Mat.* **21**, 395502 (2009). https://doi.org/10.1088/0953-8984/21/39/395502

48 *QUANTUM ESPRESSO*, <https://www.quantum-espresso.org/> (2023).

49 *Wannier90*, <https://wannier.org> (2024).

50 *Python Tight Binding (PythTB)*, <http://www.physics.rutgers.edu/pythtb/> (2023).

51 CrysAlisPro v. 1.172.43.143a (Oxford Diffraction/Agilent Technologies UK Ltd., Oxford, UK, 2024).

52 Groom, C. R., Bruno, I. J., Lightfoot, M. P. & Ward, S. C. The Cambridge Structural Database. *Acta Crystallogr B* **72**, 171–179 (2016). https://doi.org/10.1107/S2052520616003954

53 Wang, S. K., Zhu, J. M., Blackwell, R. & Fischer, F. R. Automated Tip Conditioning for Scanning Tunneling Spectroscopy. *J Phys Chem A* **125**, 1384-1390 (2021). https://doi.org/10.1021/acs.jpca.0c10731

54 Giessibl, F. J. Advances in atomic force microscopy. *Rev Mod Phys* **75**, 949–983 (2003). https://doi.org/DOI 10.1103/RevModPhys.75.949

55 Bartels, L., Meyer, G. & Rieder, K. H. Controlled vertical manipulation of single CO molecules with the scanning tunneling microscope: A route to chemical contrast. *Applied Physics Letters* **71**, 213–215 (1997). https://doi.org/10.1063/1.119503

56 Mohn, F., Schuler, B., Gross, L. & Meyer, G. Different tips for high-resolution atomic force microscopy and scanning tunneling microscopy of single molecules. *Applied Physics Letters* **102**, 073109 (2013). https://doi.org/10.1063/1.4793200

57 Necas, D. & Klapetek, P. Gwyddion: an open-source software for SPM data analysis. *Cent Eur J Phys* **10**, 181–188 (2012). https://doi.org/10.2478/s11534-011-0096-2

58 Weigend, F. Accurate Coulomb-fitting basis sets for H to Rn. *Phys Chem Chem Phys* **8**, 1057–1065 (2006). https://doi.org/10.1039/b515623h

59 Weigend, F. Hartree-Fock exchange fitting basis sets for H to Rn. *J Comput Chem* **29**, 167–175 (2008). https://doi.org/10.1002/jcc.20702

60 Kollmar, C., Sivalingam, K., Guo, Y. & Neese, F. An efficient implementation of the NEVPT2 and CASPT2 methods avoiding higher-order density matrices. *J Chem Phys* **155**, 234104 (2021). https://doi.org/10.1063/5.0072129

61 Riplinger, C. & Neese, F. An efficient and near linear scaling pair natural orbital based local coupled cluster method. *J Chem Phys* **138**, 034106 (2013). https://doi.org/10.1063/1.4773581

62  Vydrov, O. A. & Van Voorhis, T. Nonlocal van der Waals density functional: the simpler the better. *J Chem Phys* **133**, 244103 (2010). https://doi.org/10.1063/1.3521275

63  Hujo, W. & Grimme, S. Performance of the van der Waals Density Functional VV10 and (hybrid)GGA Variants for Thermochemistry and Noncovalent Interactions. *J Chem Theory Comput* **7**, 3866–3871 (2011). https://doi.org/10.1021/ct200644w

64  Mardirossian, N. & Head-Gordon, M. ωB97M-V: A combinatorially optimized, range-separated hybrid, meta-GGA density functional with VV10 nonlocal correlation. *J Chem Phys* **144**, 214110 (2016). https://doi.org/10.1063/1.4952647

65  Ekstrom, U., Visscher, L., Bast, R., Thorvaldsen, A. J. & Ruud, K. Arbitrary-Order Density Functional Response Theory from Automatic Differentiation. *J Chem Theory Comput* **6**, 1971–1980 (2010). https://doi.org/10.1021/ct100117s

66  Yanai, T., Tew, D. P. & Handy, N. C. A new hybrid exchange-correlation functional using the Coulomb-attenuating method (CAM-B3LYP). *Chem Phys Lett* **393**, 51-57 (2004). https://doi.org/10.1016/j.cplett.2004.06.011

67  Chai, J. D. & Head-Gordon, M. Systematic optimization of long-range corrected hybrid density functionals. *J Chem Phys* **128**, 084106 (2008). https://doi.org/10.1063/1.2834918

68  Heyd, J., Scuseria, G. E. & Ernzerhof, M. Hybrid functionals based on a screened Coulomb potential. *J Chem Phys* **118**, 8207–8215 (2003). https://doi.org/10.1063/1.1564060

69  Pizzi, G. *et al.* Wannier90 as a community code: new features and applications. *J Phys-Condens Mat* **32**, 165902 (2020). https://doi.org/10.1088/1361-648X/ab51ff

**Acknowledgements**

This work was primarily supported by: the US Department of Energy (DOE), Office of Science, Basic Energy Sciences (BES), Materials Sciences and Engineering Division under contract DEAC02-05-CH11231 (Nanomachine program KC1203) (molecular design, on-surface growth, tight-binding analyses), contract DE-SC0023105 (molecular precursor synthesis), and the National Energy Research Scientific

Computing Center (NERSC), a US DOE Office of Science User Facility operated under contract DE-AC02-05CH11231 award NERSC DDR-ERCAP0032308; the Office of Naval Research (ONR) under contract N00014-24-1-2134 (STM imaging and STS characterization) and N00014-19-1-2503 (STM instrumentation); the National Science Foundation under contract DMR-2325410 (DFT simulations), TACC Frontera, and ACCESS resources at the NICS; the Heising-Simons Faculty Fellows Program at UC Berkeley (F.R.F.). The CoC-MGCF is supported in part by National Institutes of Health (NIH) contract S10OD034382; the CoC-NMR is supported in part by NIH contract S10OD024998; the CoC-X-ray facility is supported in part by NIH contract S10-RR027172.

## Author Contributions

Y.Y, F.L., B.Q., and F.R.F. initiated and conceived the research, F.L, and B.Q. designed, synthesized, and characterized the molecular precursors, Y.Y. performed on-surface synthesis and STM characterization and analysis, W.T., B.Q., S.G.L., and F.R.F performed DFT and GW calculations as well as theoretical analyses, H.X.W. assisted with data interpretation, S.G.L. and F.R.F. secured funding and supervised all aspects of the project, B.Q., Y.Y., F.L., W.T, S.G.L. and F.R.F. wrote the manuscript. All authors contributed to the scientific discussion.

## Author Information

The Reprints and permissions information is available at www.XXX/. The authors declare no competing financial interests. Correspondence and requests for materials should be addressed to ffischer@berkeley.edu or sglouie@berkeley.edu.

**Main Figures**

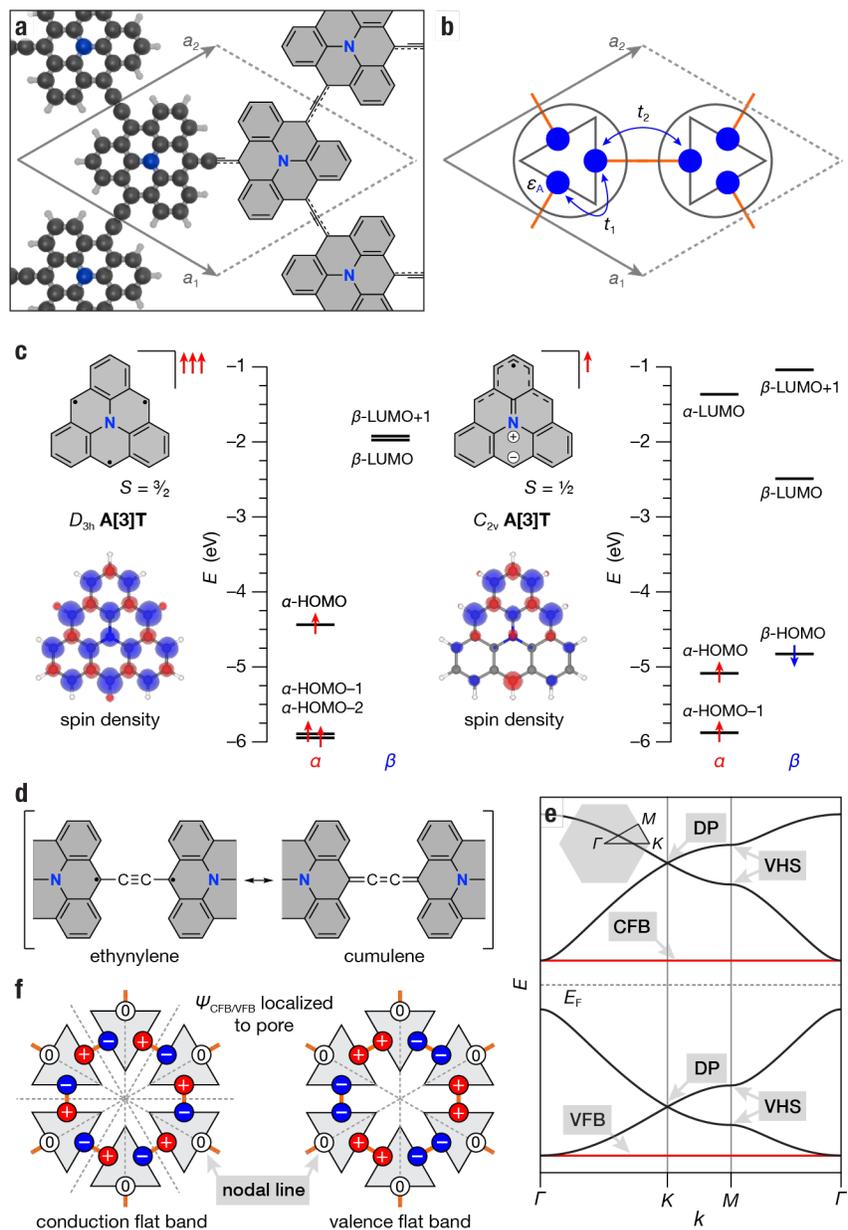

**Figure 1 | Orbital engineering of low-dimensional A[3]TCOFs. a,** Molecular model of the A[3]TCOFs assembled from aza-[3]triangulene core and ethynylene linkers. **b,** Schematic representation of a NN TB model including the intra- and inter-A[3]T electron hopping parameters $t_1$, and $t_2$, and the onsite energy $\varepsilon_A$ for a diatomic Kagome lattice. **c,** Molecular structure and DFT calculated orbital energies (M06-2X/def2-TZVP) for $D_{3h}$ and $C_{2v}$ symmetric A[3]T. Projections of the local spin density onto the structure of a $D_{3h}$ and $C_{2v}$ symmetric A[3]T. **d,** Schematic representation of the ethynylene-cumulene resonance structure that

contributes to the stabilization of the $D_{3h}$ symmetric ground state of aza-[3]triangulene cores in 2D A[3]TCOFs. **e,** Model band structure of a Type-II diatomic Kagome lattice. Phase-frustration-induced Kagome conduction (CFB) and valence flat bands (VFB), Dirac points (DP) and Van Hove singularities (VHS) are highlighted by arrows. **f,** Schematic representation of the destructive wave function interference that gives rise to the CFB and VFB. The alternating sign of the MLWFs is represented by filled circles (red, positive; blue, negative). Orbital nodes resulting from destructive wave function interference are indicated by dashed lines and white circles (contributions from MLWF located at the position of nodal planes is zero).

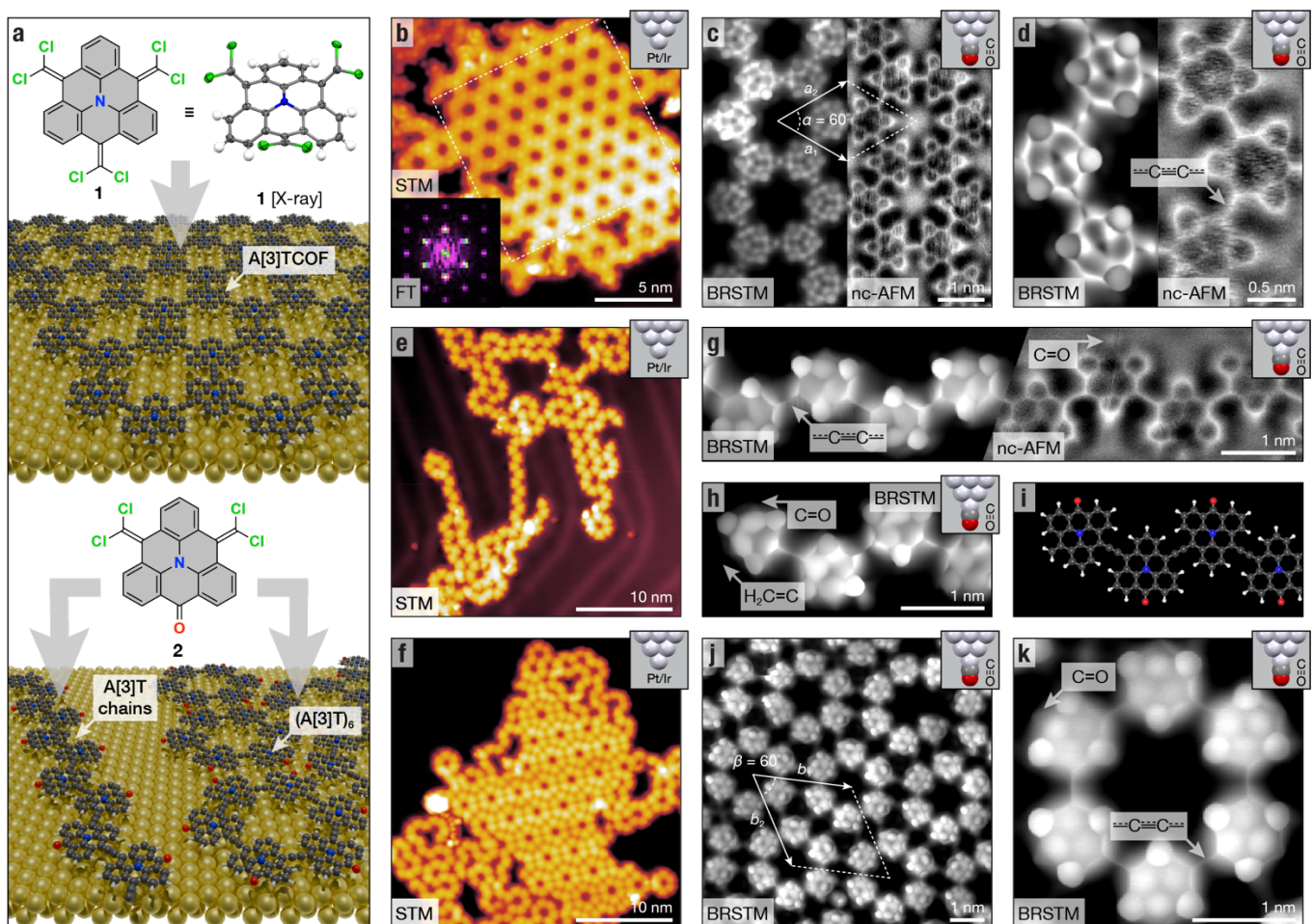

**Figure 2 | Bottom-up synthesis of A[3]TCOFs. a,** Schematic representation of the on-surface synthesis of A[3]TCOFs (top) and linear (cyclic) A[3]T chains (bottom) from molecular precursors **1** and **2**, respectively. Single X-ray crystal structure of **1**. Thermal ellipsoids are drawn at the 90% probability level. Colour coding: C (grey), N (blue), Cl (green). H-atoms are placed at the calculated positions. **b,** STM topographic image of a locally ordered island of A[3]TCOF formed by annealing a monolayer of **1** on Au(111) to $T$ = 523 K ($V_s$ = –600 mV, $I_t$ = 20 pA). Inset shows the Fourier transform (FT) of the Kagome lattice enclosed by a white dashed line. **c,** Constant-height BRSTM image (left) and nc-AFM image (right) of a representative segment of the *p6mm* lattice of A[3]TCOF ($V_s$ = –10 mV, $V_{ac}$ = 10 mV, $f$ = 455 Hz, CO-functionalized tip). Unit cell is highlighted by a white dashed line. **d,** Constant-height BRSTM image (left) and nc-AFM image (right) of a representative segment of the *p6mm* lattice of A[3]TCOF. Arrows highlight position of cumulene/ethynylene linkers ($V_s$ = –10 mV, $V_{ac}$ = 10 mV, $f$ = 455 Hz, CO-functionalized tip). **e,**

STM topographic image of linear A[3]T chains formed by annealing a monolayer of **2** on Au(111) to $T$ = 523 K ($V_s$ = –600 mV, $I_t$ = 20 pA). **f,** STM topographic image of a self-assembled island of cyclic A[3]T chains formed by annealing a monolayer of **2** on Au(111) to $T$ = 523 K ($V_s$ = –600 mV, $I_t$ = 20 pA). **g,** Constant-height BRSTM image (left) and nc-AFM image (right) of a segment of a linear A[3]T chain ($V_s$ = –10 mV, $V_{ac}$ = 10 mV, $f$ = 455 Hz, CO-functionalized tip) Arrows highlight position of cumulene/ethynylene linkers and carbonyl groups. **h,** Constant-height BRSTM image of an end group of a A[3]T chain ($V_s$ = –10 mV, $V_{ac}$ = 10 mV, $f$ = 455 Hz, CO-functionalized tip). Arrows highlight position of carbonyl and vinylidene end groups. **i,** Molecular model of the A[3]T end group imaged in (H). **j,** Constant-height BRSTM image of a self-assembled lattice of cyclic $(A[3]T)_6$ ($V_s$ = –10 mV, $V_{ac}$ = 10 mV, $f$ = 455 Hz, CO-functionalized tip). Unit cell is highlighted by a white dashed line. **k,** Constant-height BRSTM image of a single six-membered ring of $(A[3]T)_6$ ($V_s$ = –10 mV, $V_{ac}$ = 20 mV, $f$ = 455 Hz, CO-functionalized tip). Arrows highlight position of cumulene/ethynylene linkers and carbonyl groups. All STM experiments were performed at $T$ = 4.5 K.

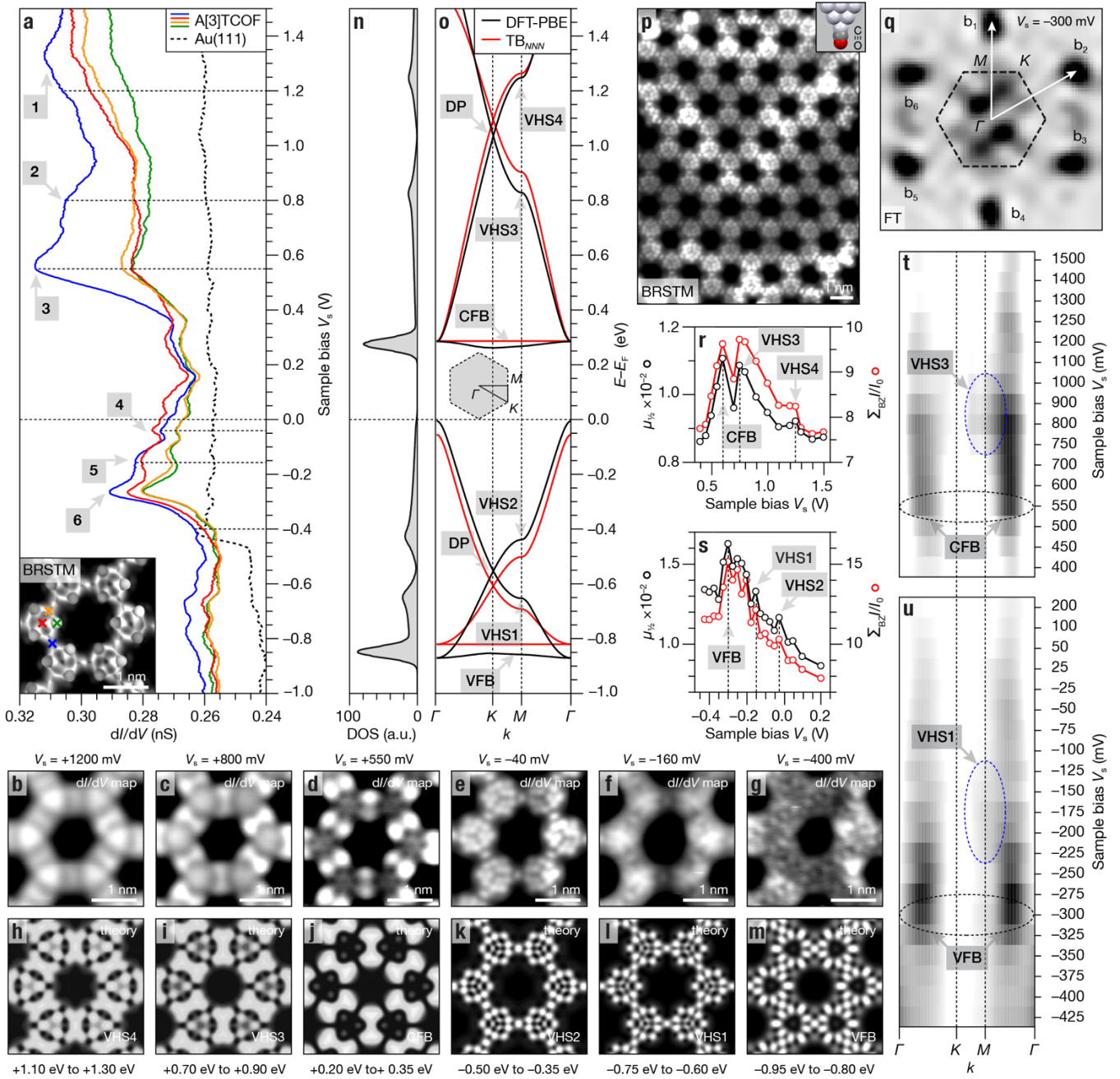

**Figure 3 | Electronic structure of A[3]TCOF Kagome lattices. a,** d$I$/d$V$ point spectroscopy of A[3]TCOF on Au(111) recorded at the positions marked in the BRSTM inset (blue, red, orange, and green crosses; spectroscopy: $V_{ac}$ = 10 mV, $f$ = 455 Hz; constant-height imaging: $V_s$ = −10 mV, $V_{ac}$ = 10 mV, $f$ = 455 Hz, CO-functionalized tip). Arrows highlight the position of *features* 1–6; dashed lines indicate the sample bias for d$I$/d$V$ maps in (b–g). **b–g,** Constant-height d$I$/d$V$ maps recorded at a sample voltage bias of $V_s$ = +1200 mV, $V_s$ = +800 mV, $V_s$ = +550 mV, $V_s$ = −40 mV, $V_s$ = −160 mV, and $V_s$ = −450 mV, respectively ($V_{ac}$ = 10

mV, $f$ = 455 Hz, CO-functionalized tip). **h–m,** DFT-PBE LDOS maps above a freestanding A[3]TCOF at energies corresponding to VHS4, VHS3, CFB, VHS2, VHS1, and VFB calculated using the model for a $p$-wave ($E$–$E_F$ > 0 eV) or $s$-wave ($E$–$E_F$ < 0 eV) STM tip. LDOS maps are sampled at a height of 3 Å above the atomic plane of the A[3]TCOF. **n,** DFT-PBE DOS for a freestanding A[3]TCOF (spectrum broadened by 27 meV Gaussian). **o,** DFT-PBE band structure for a freestanding A[3]TCOF Kagome lattice. **p,** Constant-height BRSTM image recorded at a sample voltage bias of $V_s$ = –10 mV on a locally ordered 5 × 8 lattice of A[3]TCOF showing QPI patterns ($V_{ac}$ = 10 mV, $f$ = 455 Hz, CO-functionalized tip). **q,** FT of constant-height d$I$/d$V$ map recorded at $V_s$ = +550 mV ($V_{ac}$ = 10 mV, $f$ = 455 Hz, CO-functionalized tip). Dashed hexagon encloses the first Brillouin zone ($^{1st}$BZ). **r–s,** Plot of the FT signal intensity ($\sum_{BZ} \frac{I}{I_0}$, red circles) and median intensity ($\mu_{\frac{1}{2}}$, black circles) integrated over the 1$^{st}$ BZ. **t–u,** Plot of the FT amplitude as a function of wavevector $\boldsymbol{k}$ averaged over all $\Gamma$–$K$–$M$–$\Gamma$ paths at the indicated sample biases. All STM experiments were performed at $T$ = 4.5 K.

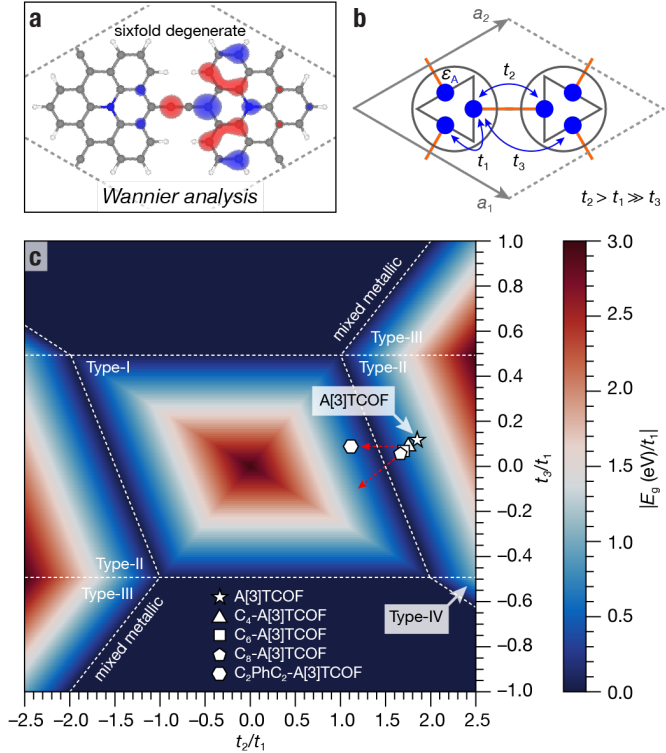

**Figure 4 | Electronic structure of A[3]TCOF Kagome lattices. a,** Projection of the sixfold degenerate MLWF associated with the cumulene/ethynylene linker onto the structure of the A[3]T core (Site A). **b,** Schematic representation of intra- and inter-A[3]T electron hopping parameters $t_1$, $t_2$, and $t_3$ along with onsite energy $\varepsilon_A$ for a diatomic Kagome lattice. **c,** Phase diagram for the TB Kagome band structure as a function of $t_2$ and $t_3$ (in units of $t_1$). Colour gradient indicates the size of the band gap in $E_g$ (eV)/$t_1$; Type-I, Type-II, Type-III, Type-IV, and mixed metallic Kagome band structures are separated by dashed lines. A star marks the position of the A[3]TCOF. Filled polygons mark the position of A[3]T lattices featuring buta-1,3-diyne (triangle), hexa-1,3,5-triyne (square), octa-1,3,5,7-tetrayne (pentagon), and 1,4-diethynylbenzene (hexagon) linkers derived by fitting a truncated NNN TB Hamiltonian to DFT-PBE calculations.

**Data Availability**

The DFT code and pseudopotentials can be downloaded from the Quantum Espresso website [47,48]. For this study, we used version 6.7 for the DFT-PBE calculations. The Wannier90 code can be downloaded from the Wannier90 website [49]. The tight-binding simulation code can be downloaded from the PythTB website [50]. For this study, we used version 1.8.0 for the tight-binding fitting. All data are available in the main text or the Supplementary Materials. X-ray crystal structure data was refined using the CrysAlis[Pro] [51] software package and can be accesses through the Cambridge Structural Database (CSD) CCDC 2416619 [52].

**Methods**

**Precursor Synthesis and A[3]TCOF Growth.** Full details of the synthesis and characterization of **1** and **2** are given in the Supplementary Information. A[3]TCOFs were grown on Au(111)/mica films under UHV conditions. Atomically clean Au(111) substrates were prepared using alternating $Ar^+$ sputtering and annealing cycles under UHV conditions. Sub-monolayer coverages of precursor **1** or **2** were prepared by UHV deposition using a commercial MBE cell evaporator operating at crucible temperatures of 463 K for 5 min with the substrate held at 297 K. Substrate temperatures were then gradually ramped (~3 K min$^{-1}$) to $T$ = 523 K and held for 30 min to induce the growth of 2D A[3]TCOF, linear A[3]T chains, and cyclic $(A[3]T)_6$, respectively.

**STM Measurements.** All STM experiments were performed using a commercial OMICRON LT-STM held at $T$ = 4.5 K using PtIr STM tips. STM tips were optimized for scanning tunnelling spectroscopy using an automated tip conditioning program[53]. d$I$/d$V$ spectra were recorded with CO-functionalized STM tips using a lock-in amplifier with a modulation frequency of 455 Hz and a modulation amplitude of $V_{RMS}$ = 10–20 mV. d$I$/d$V$ point spectra were recorded under open feed-back loop conditions. d$I$/d$V$ maps were collected under constant height conditions. BRSTM images were obtained by mapping the current signal collected during a constant-height scanning at $V_s$ = –10 mV. Each peak position is based on an average of 3 to 5 spectra, all of which were first calibrated to the Au(111) Shockley surface state. nc-AFM measurements were performed in a qPlus-equipped[54] commercial Omicron LT-STM/AFM in constant-height mode using CO-functionalized tips[55,56] (resonance frequency, $f$ = 27.64 kHz; nominal spring constant, $k$ = 1,800 N m$^{-1}$; Q-Factor, $Q$ = 44,000; oscillation amplitude, $A$ = 125 pm; bias voltage $V_s$ = –10 mV). All images were processed using the Gwyddion software package[57].

**Fourier transform analysis of d$I$/d$V$ maps.** The discrete Fourier transform (FT) was applied to each d$I$/d$V$ map after zeroing their means. We initially considered the magnitude $I(\boldsymbol{k})$ (rather than the phase) component of each FT image, which carries information on the magnitude of contributions of states with different wavevectors to the overall differential conductance signal. To obtain a fair basis for comparing between maps obtained at different biases, we normalized the intensities in each FT image by the integrated intensity of the six peaks corresponding to the first-order reciprocal lattice points $\{\boldsymbol{K}_i\}$, which correspond to the period $a$ of the unit cell, and best represent the overall differential conductance of each map:

$$\frac{I}{I_0} = \frac{I}{\sum_{|\boldsymbol{K}_i - \boldsymbol{k}| < \Delta \boldsymbol{k}} I}$$

where $\Delta \boldsymbol{k}$ denotes the broadening of the peaks at $\{\boldsymbol{K}_i\}$ in our discrete FT image. Then, we focused our analysis on the strength of long-wavelength contributions, which are contained in the first Brillouin zone (1BZ). From the points in the 1BZ of the normalized images, two statistics were calculated: the total normalized intensity $\sum_{BZ} \frac{I}{I_0}$ and median normalized intensity $\mu_{\frac{1}{2}}$ (half the pixels in the 1BZ have normalized intensity higher/lower than $\mu_{\frac{1}{2}}$). For visual inspection, the raw FT images were interpolated by Lanczos resampling.

**Simulation of A[3]T core and A[3]TCOF dimer.** All density functional theory computations were performed using the Orca software package[27]. Gas-phase geometry optimizations were performed using the M06-2X exchange-correlation functional and the def2-SVP basis set[25,28]. Single-point energy calculations were performed using the M06-2X, CAM-B3LYP, ωB97X-D3, ωB97M-V functionals with a def2-TZVP and AuxJ basis set[58], the CASSCF-NEVPT2 with a def2-TZVP and AuxJK basis set[59-61], and DLPNO-CCSD(T) with def2-TZVPP, AuxJ and AuxC basis set at the relaxed geometry for the ground state energy[62-67]. Orbital information was edited using Multiwfn software[26,29].

**Simulation of A[3]TCOF, liner A[3]T chains, and cyclic (A[3]T)$_6$.** First-principles DFT calculations using Perdew-Burke-Ernzerhof (PBE) functional[39] and Heyd-Scuseria-Ernzerhof (HSE06) functional[68] were performed using the Quantum Espresso package[47,48]. We used norm-conserving pseudopotentials with a 100 Ry energy cut-off and 0.002 Ry Gaussian broadening. A supercell geometry was employed, with a 10 Å vacuum spacing placed in all non-periodic directions to prevent interaction between replicas. For the self-consistent field calculation, a $12 \times 12 \times 1$ **k**-grid was used for the 2D structure and a $20 \times 1 \times 1$ **k**-grid was used for the 1D structure. The structures were first fully relaxed until all components of the force were smaller than 0.01 eV Å$^{-1}$ using PBE functionals. The PBE band structure calculations were performed using 40 **k**-points along the periodic directions. The HSE band structure was interpolated using Wannier90 package[49,69]. We used the Wannier90 package for acquiring maximally localized Wannier functions (MLWFs) from DFT-PBE results and used the PythTB[50] package to fit tight-binding parameters.

**Supplementary Information**

Supplementary Information contains detailed synthetic procedures and characterization of precursors **1** and **2**; Electronic structure of linear A[3]T chains and cyclic (A[3]T)$_6$; Bonding analysis for linear A[3]T chains and cyclic (A[3]T)$_6$; Supplementary Information Figures S1 to S9; X-ray Crystal Structure Data of **1**; Supplementary Information Tables S1 to S6

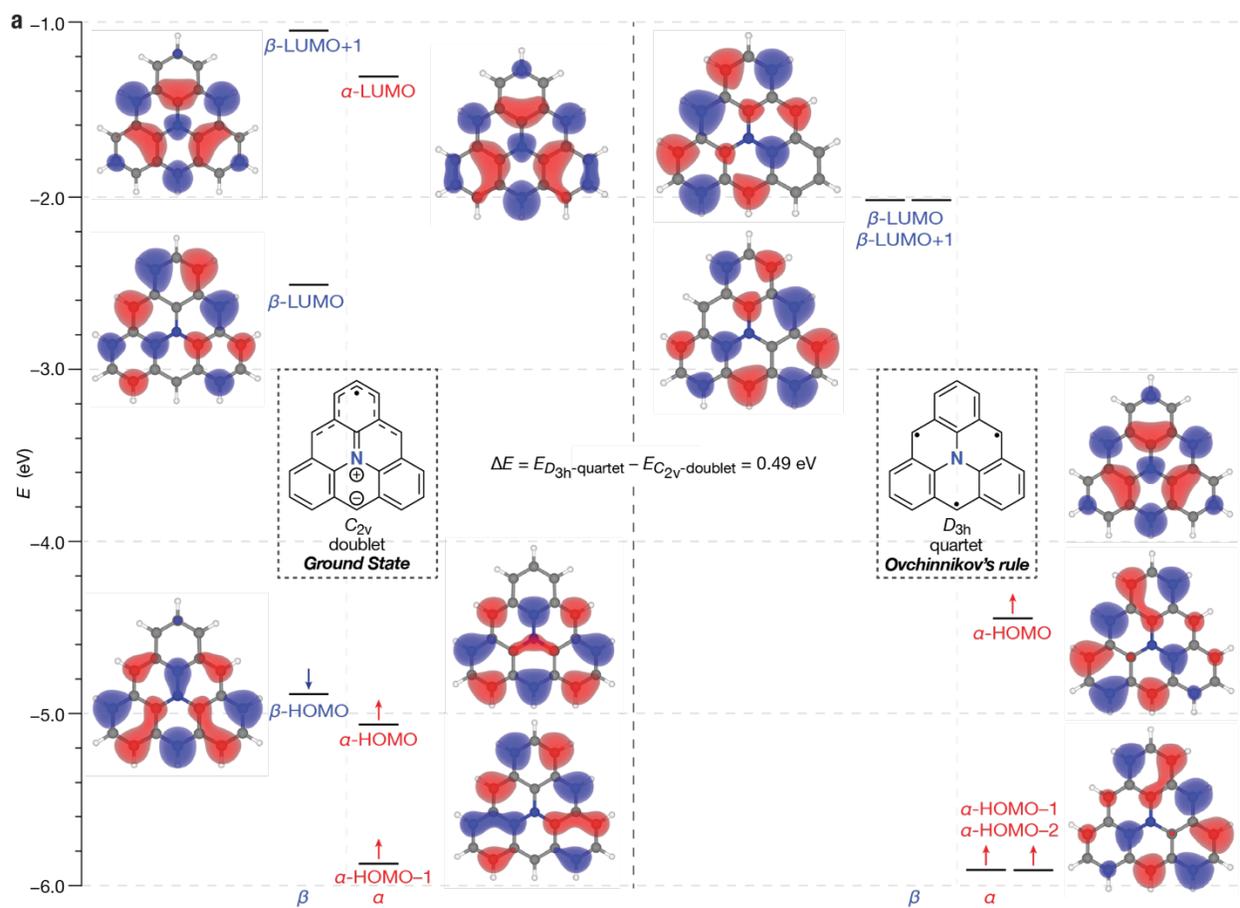

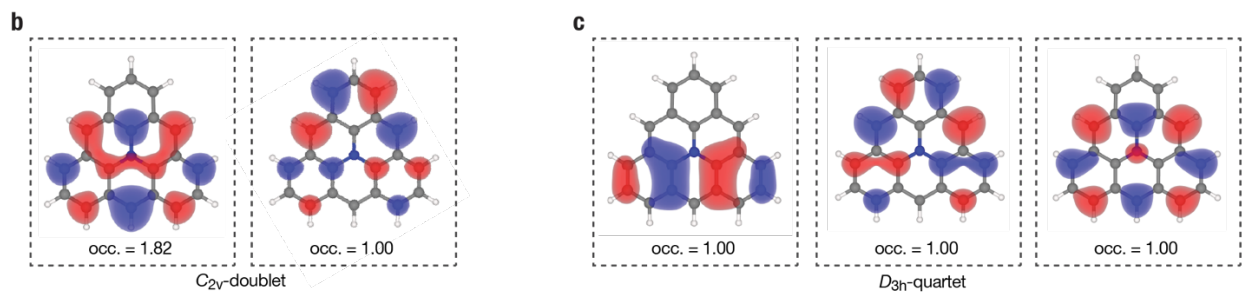

**d**

| energy (eV) | $C_{2v}$– doublet | $D_{3h}$– quartet | $\Delta E_{D_{3h}\text{-quartet} - C_{2v}\text{-doublet}}$ |
|---|---|---|---|
| M06-2X | −23460.53 | −25460.04 | 0.49 |
| CAM-B3LYP | −23456.72 | −23456.50 | 0.22 |
| ωB97M-V | −23461.59 | −23461.32 | 0.27 |
| ωB97X-D3 | −25462.79 | −25462.60 | 0.19 |
| DLPNO-CCSD(T) | −25417.79 | −25417.59 | 0.20 |

**Extended Data Figure ED1 | Ab initio calculation of high-spin $D_{3h}$ and low-spin $C_{2v}$ symmetric A[3]T.**

**a,** Molecular structure and DFT calculated orbitals for high-spin $D_{3h}$ and low-spin $C_{2v}$ symmetric A[3]T

(M06-2X/def2-TZVP). **b,** Selected natural orbitals in CAS(5,5) for low-spin $C_{2v}$ symmetric A[3]T with different occupations. **c,** Selected natural orbitals in CAS(5,6) for high-spin $D_{3h}$ symmetric A[3]T with different occupation. **d,** Optimized energy for high-spin $D_{3h}$ and low-spin $C_{2v}$ symmetric A[3]T at different levels of theory.

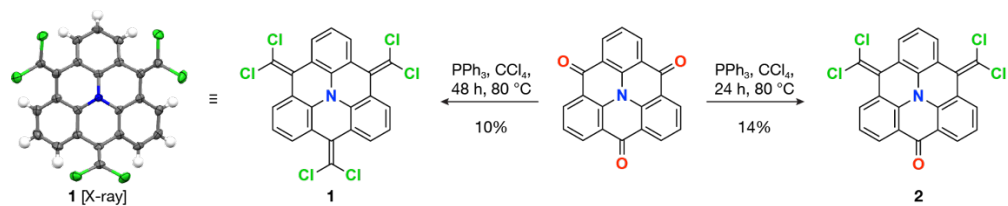

**Extended Data Figure ED2 | Synthesis of molecular precursors for A[3]TCOF and linear A[3]T chains and cyclic (A[3]T)₆.** Synthesis of 4,8,12-tris(dichloromethylene)-8,12-dihydro-4*H*-benzo[9,1]quinolizino[3,4,5,6,7-*defg*]acri-dine (**1**) and 8,12-bis(dichloromethylene)-8,12-dihydro-4*H*-benzo[9,1]quinolizino [3,4,5,6,7-*defg*]acridin-4-one (**2**) from 4*H*-benzo[9,1]quinolizino[3,4,5,6,7-*defg*]acridine-4,8,12-trione.

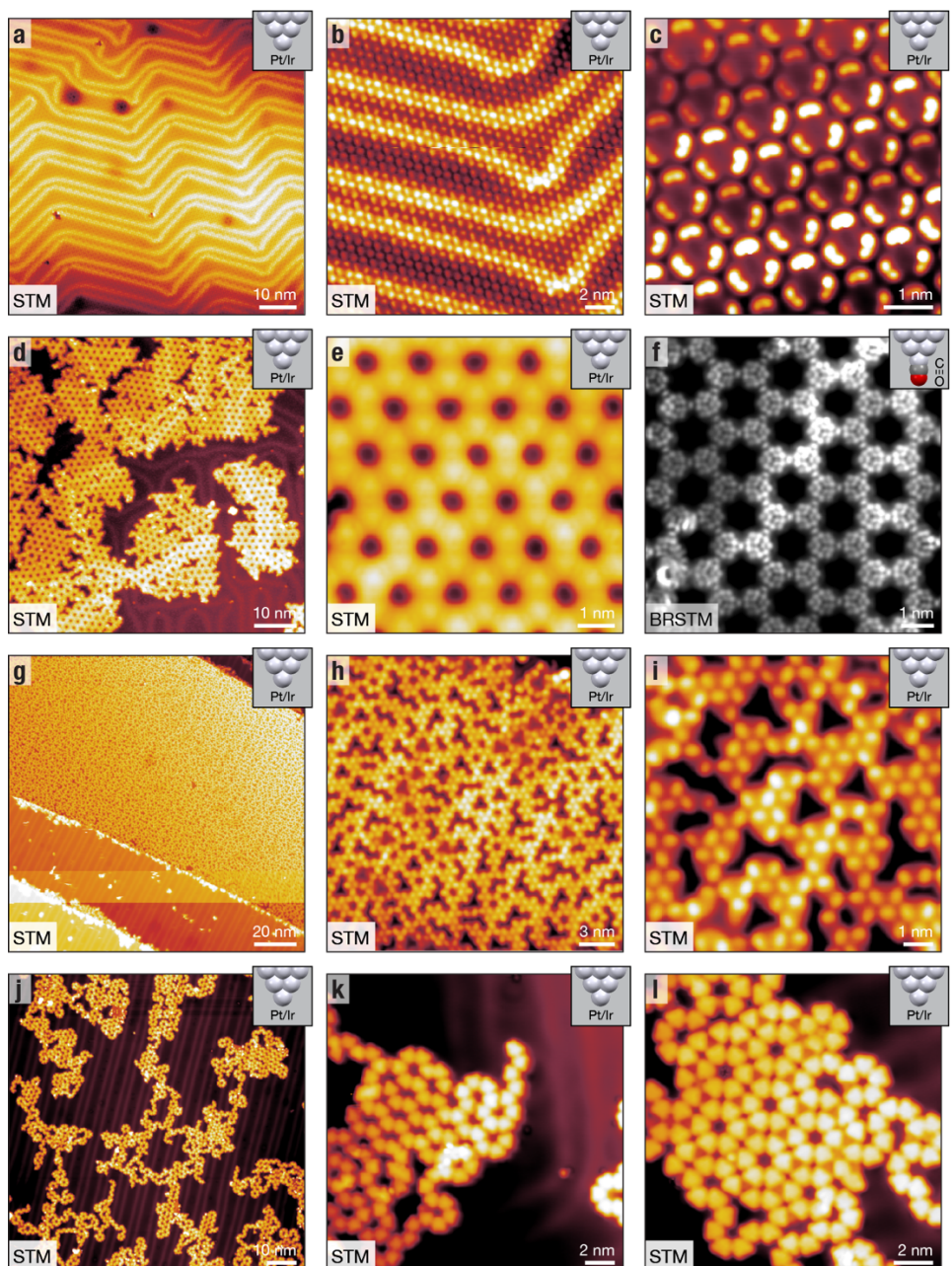

**Extended Data Figure ED3 | On-surface synthesis of A[3]TCOF, linear A[3]T chains, and cyclic (A[3]T)₆. a–c,** STM topographic image of dense-pack self-assembly of **1** after sublimation on Au(111). (a, $V_s = -600$ mV, $I_t = 20$ pA; b, $V_s = -200$ mV, $I_t = 30$ pA; c, $V_s = -100$ mV, $I_t = 20$ pA). **d–e,** STM topographic image of polymerized **1** at 523 K. (D, $V_s = -1000$ mV, $I_t = 30$ pA; E, $V_s = -400$ mV, $I_t = 30$ pA). **f,** Constant-height BRSTM image ($V_s = -10$ mV, CO-functionalized tip) of polymerized **1**, showing the formation of the A[3]TCOF. **g–i,** STM topographic image of dense-pack self-assembly of **2** after sublimation on Au(111). (g, $V_s = -600$ mV, $I_t = 20$ pA; h, $V_s = -200$ mV, $I_t = 30$ pA; i, $V_s = -100$ mV, $I_t = 20$ pA). **j–l,** STM

topographic image of polymerized **2** at 523 K, showing the formation of the linear A[3]T chains and cyclic (A[3]T)$_6$. (j, $V_s$ = –1000 mV, $I_t$ = 30 pA; k, $V_s$ = –400 mV, $I_t$ = 30 pA; l, $V_s$ = –200 mV, $I_t$ = 50 pA). All STM experiments were performed at $T$ = 4.5 K.

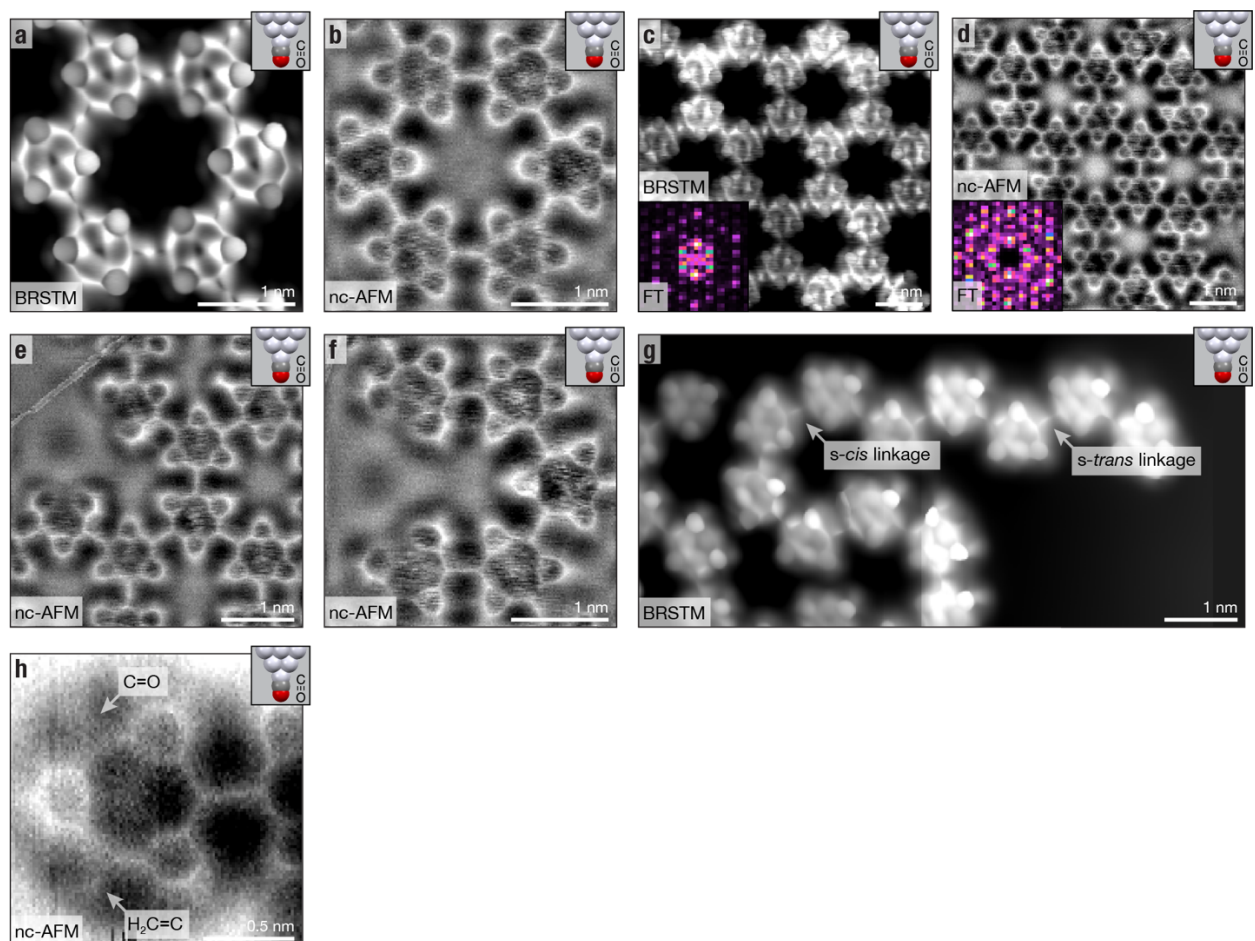

**Extended Data Figure ED4 | BRSTM and nc-AFM images of A[3]TCOF and linear A[3]T chains. a,**

Constant-height BRSTM image of a representative segment of the *p6mm* lattice of A[3]TCOF ($V_s = -10$

mV, $V_{ac} = 10$ mV, $f = 455$ Hz, CO-functionalized tip). **b,** nc-AFM image of a representative segment of the

*p6mm* lattice of A[3]TCOF ($V_s = -10$ mV, $f = 27.64$ kHz, CO-functionalized tip). **c,** Constant-height

BRSTM image of a representative segment of the *p6mm* lattice of A[3]TCOF ($V_s = -10$ mV, $V_{ac} = 10$ mV,

$f = 455$ Hz, CO-functionalized tip). Inset shows the Fourier transform (FT) of the electronic structure of the

Kagome lattice (*note:* clear signal intensity is observed in the first BZ). **d,** nc-AFM image of a representative segment of the *p6mm* lattice of A[3]TCOF ($V_s = -10$ mV, $f = 27.64$ kHz, CO-functionalized tip). Inset shows the Fourier transform (FT) of the atomic structure of the Kagome lattice (*note:* no signal intensity is observed in the first BZ suggesting the QPI patterns are purely electronic in nature). **e–f,** Two representative

nc-AFM images showing the edge termination of a locally ordered A[3]TCOF island ($V_s = -10$ mV, $f =$

27.64 kHz, CO-functionalized tip). **g,** Constant-height BRSTM image of a representative linear A[3]T chain featuring segments of s-*cis* and s-*trans* linked A[3]T cores ($V_s$ = –10 mV, $V_{ac}$ = 10 mV, $f$ = 455 Hz, CO-functionalized tip). **h,** nc-AFM image of a A[3]T chain end featuring the C=O and C=CH$_2$ end groups.

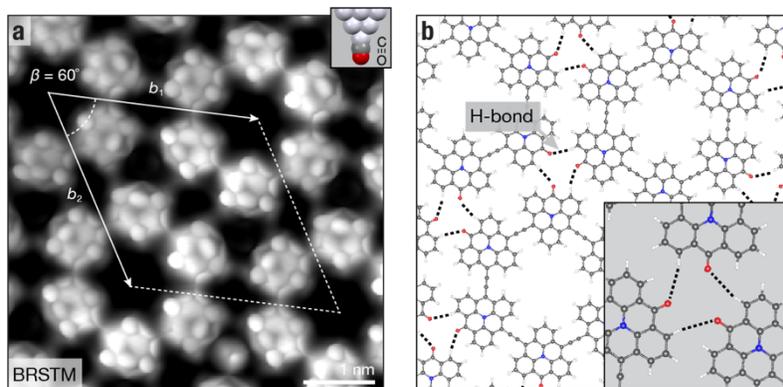

**Extended Data Figure ED5 | Model for the on-surface self-assembly of (A[3]T)$_6$ into densely packed hexagonal lattices. a,** Constant-height BRSTM image of a self-assembled lattice of cyclic (A[3]T)$_6$ ($V_s$ = −10 mV, $V_{ac}$ = 10 mV, $f$ = 455 Hz, CO-functionalized tip). Unit cell is highlighted by a white dashed line. **b,** Schematic representation of the weak C–H•••O=C hydrogen bonding pattern that directs the assembly of (A[3]T)$_6$ on the Au(111) surface.

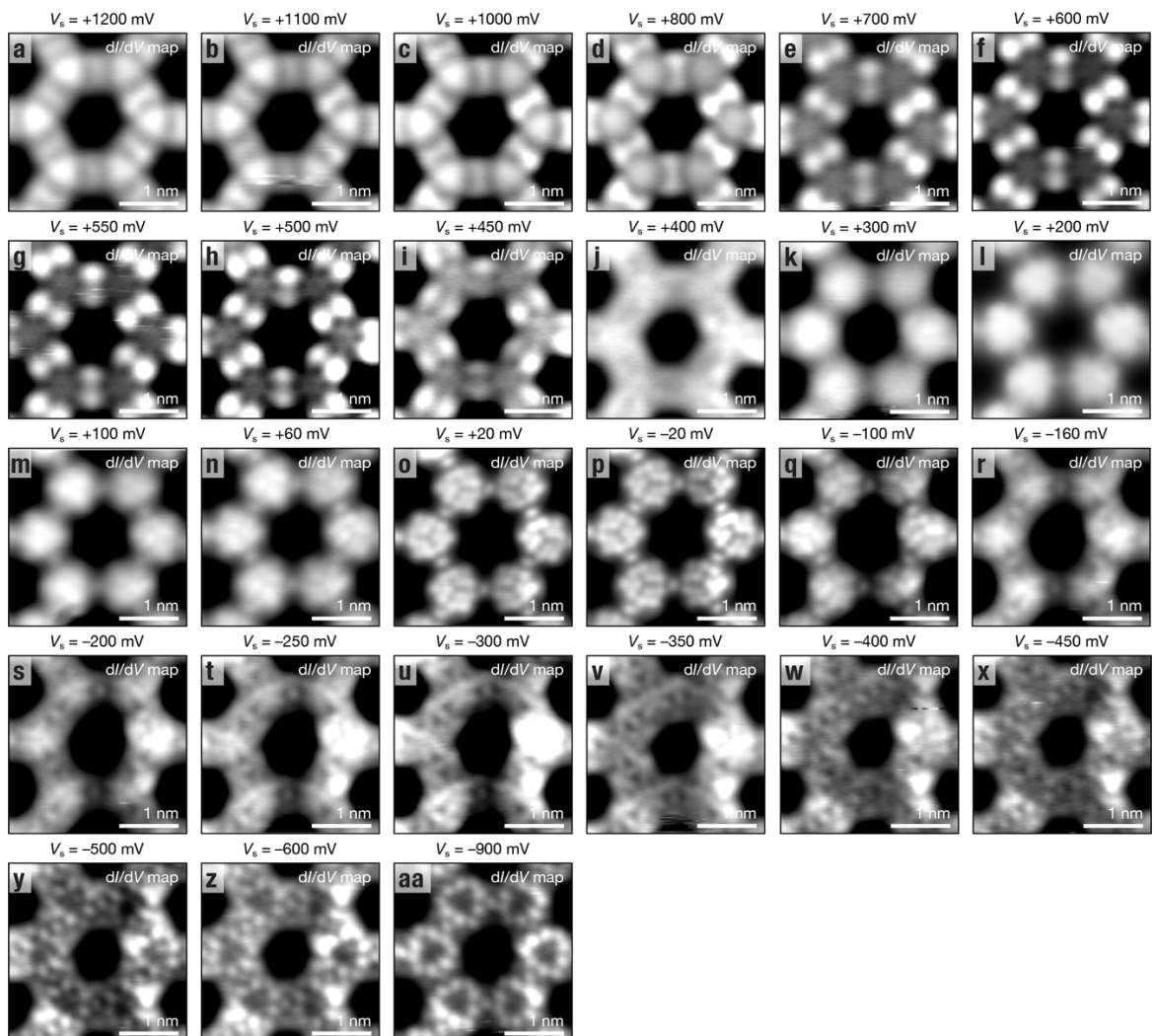

**Extended Data Figure ED6 | Electronic structure of A[3]TCOF. a–aa,** Constant-height d*I*/d*V* maps recorded at the indicated sample voltage biases on the same A[3]TCOF depicted in Figure 3 using different STM tips (*V*ac = 10–20 mV, *f* = 455 Hz, CO-functionalized tip). All STM experiments were performed at *T* = 4.5 K.

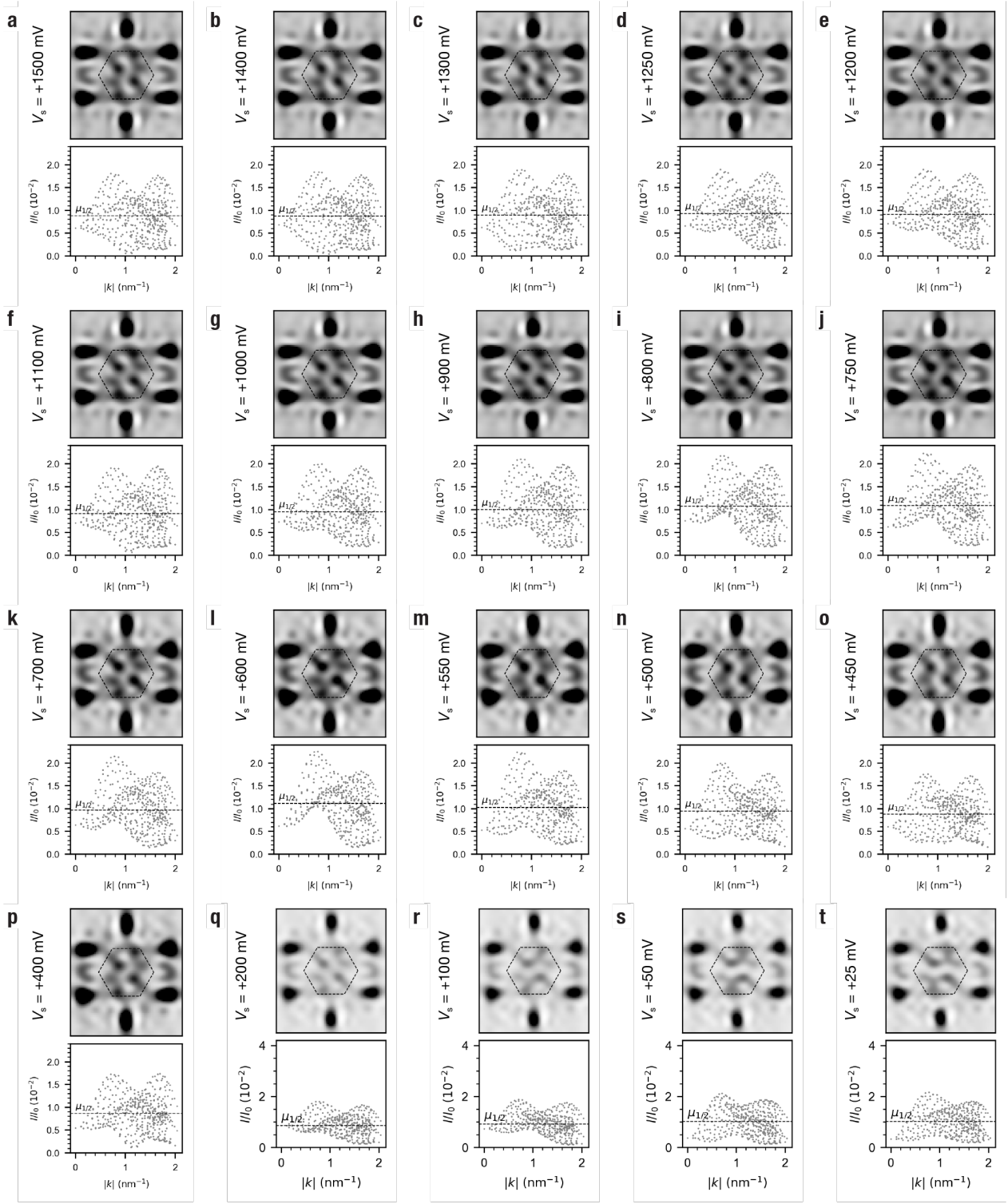

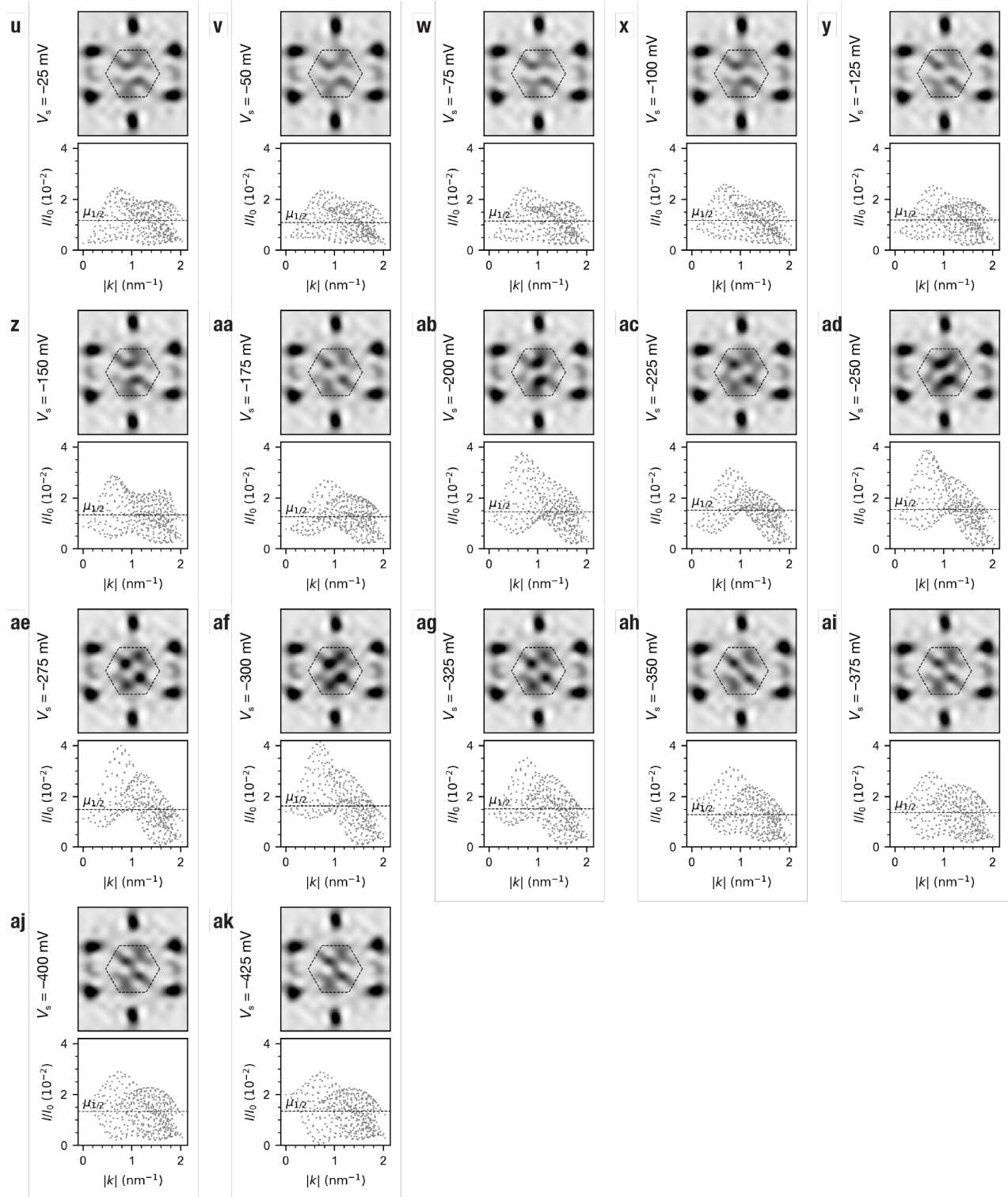

**Extended Data Figure ED7 | Momentum resolved FT-STM recorded on a 5 × 8 A[3]TCOF Kagome lattice. a–ak,** (top) FT of constant-height d$I$/d$V$ maps recorded at the indicated sample voltage biases ($V_{ac}$ = 10 mV, $f$ = 455 Hz, CO-functionalized tip). (bottom) Normalized FT signal intensity ($I$/$I_0$) as a function

of the magnitude of the momentum vector $|k|$ (with respect to the $\Gamma$ point) within the first BZ. Median normalized intensity ($\mu_{\frac{1}{2}}$) indicated as dashed line.

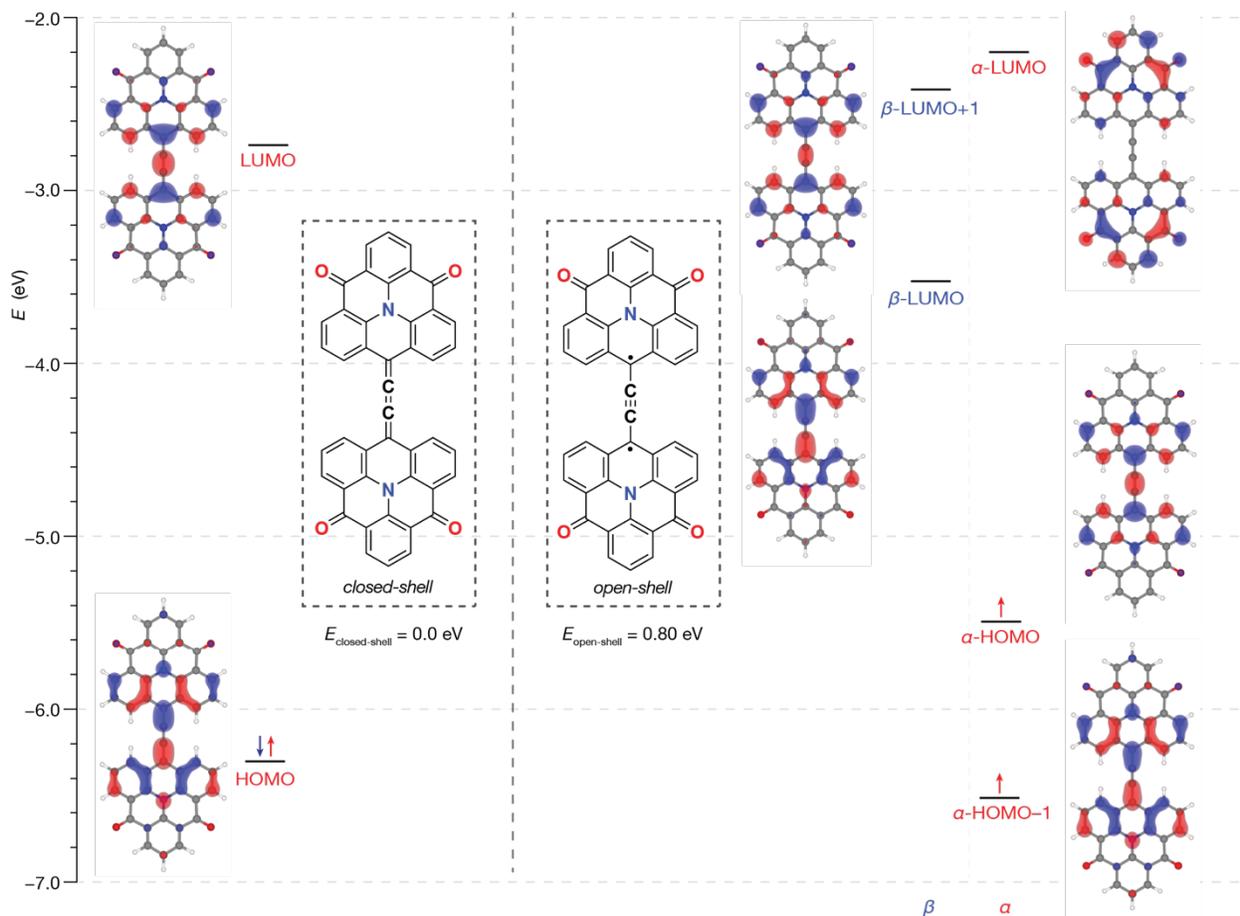

**Extended Data Figure ED8 | Molecular structure and DFT calculated orbital energies (M06-2X/def2-TZVP) for a closed-shell and high-spin aza-triangulene dimer.** The unrestricted and restricted singlet simulations show identical ground states with $<S^2> = 0$ and are 0.80 eV lower in energy than the open-shell triplet state.

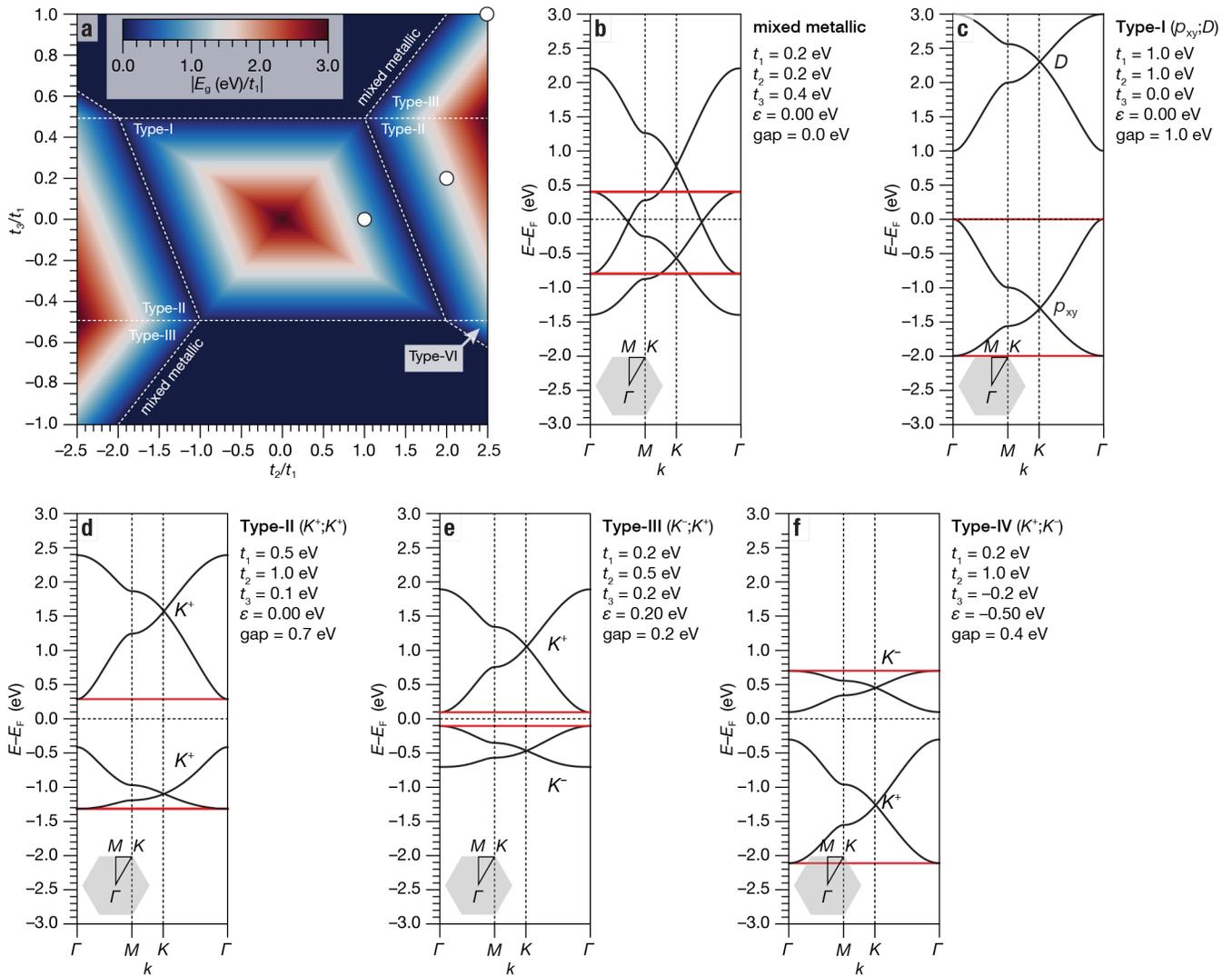

**Extended Data Figure ED9 | Typical TB band structures within the phase diagram of Kagome lattices. a,** Phase diagram for the TB Kagome band structure as a function of $t_2$ and $t_3$ (in units of $t_1$). Colour gradient indicates the size of the band gap in $|E_g \text{ (eV)}/t_1|$; Type-I ($p_{xy}$;$D$), Type-II ($K^+$;$K^+$), Type-III ($K^-$;$K^+$), Type-IV ($K^+$;$K^-$), and mixed metallic Kagome band structures are separated dotted lines. Filled circles mark the position of examples in (b–f) that fall within the depicted window. **b,** Typical TB band structure of a mixed metallic phase. Dispersive Dirac bands cross the $E_F$. **c,** Typical TB band structure of a Type-I phase. The top two Dirac bands ($D$) arise typically from $s$-orbital hoppings in a hexagonal lattice while the four band VB complex ($p_{xy}$) is usually derived from $p_{xy}$-orbital hoppings. **d,** Typical TB band structure of a Type-II phase. Pairs of two Dirac bands bordered at lower energy by a flat band ($K^+$, the sign in the exponent indicates the sign of the lattice hopping; positive the flat band is at found at lower energy; negative the flat

band is at found at higher energy). **e,** Typical TB band structure of a Type-III phase ($K^-$; $K^+$). (F) Typical TB band structure of a Type-IV phase ($K^+$; $K^-$).

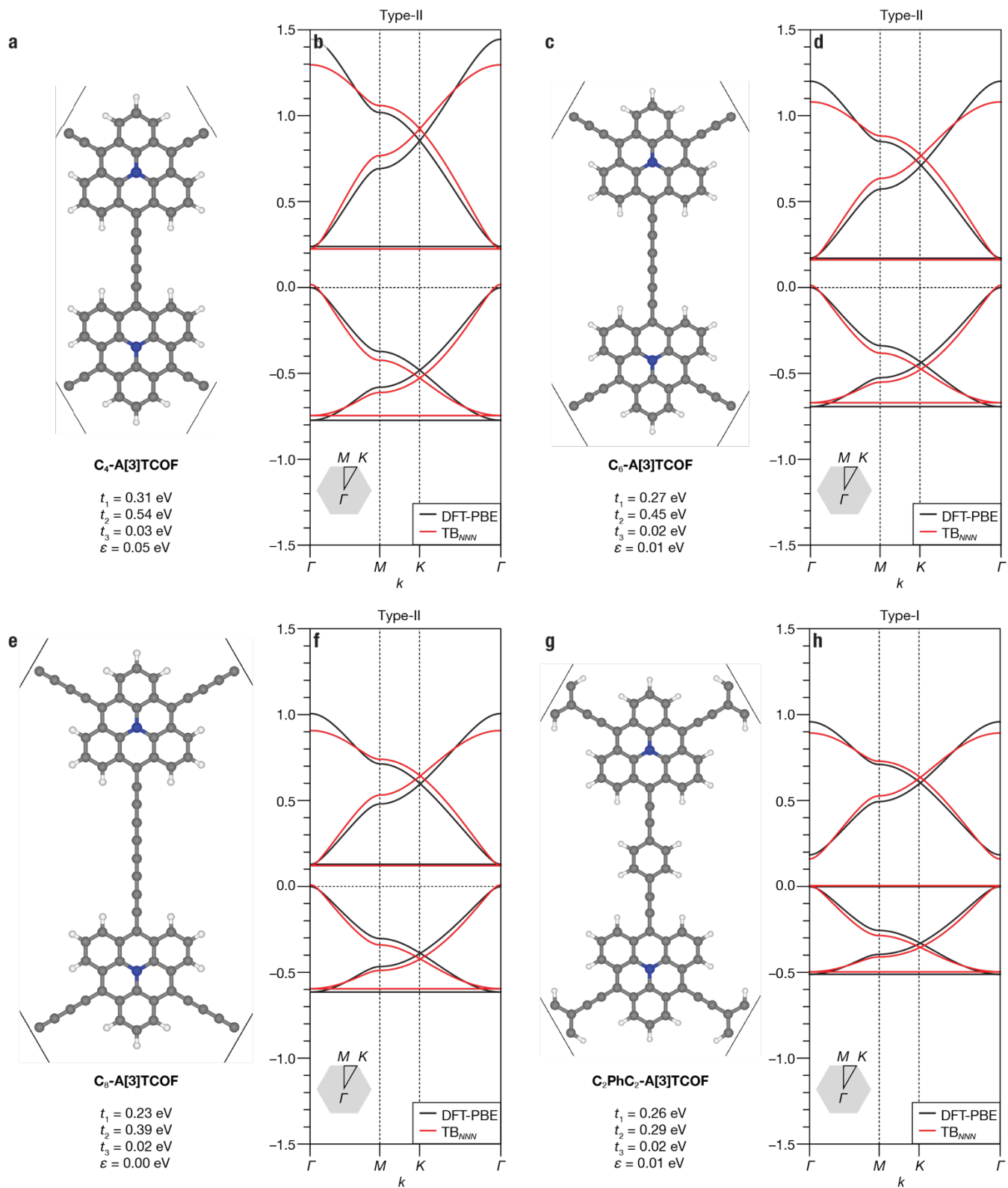

**Extended Data Figure ED10 | Electronic structure of extended A[3]TCOF featuring extended linkers.**

**a,** Molecular model of a $C_4$-A[3]TCOFs assembled from aza-[3]triangulene core and buta-1,3-diyne linkers.

**b,** DFT-PBE band structure for a freestanding $C_4$-A[3]TCOF and corresponding truncated NNN TB Hamiltonian in the Wannier basis. **c,** Molecular model of a $C_6$-A[3]TCOFs assembled from aza-[3]triangulene core and hexa-1,3,5-triyne linkers. **d,** DFT-PBE band structure for a freestanding $C_6$-A[3]TCOF and corresponding truncated NNN TB Hamiltonian in the Wannier basis. **e,** Molecular model of a $C_8$-A[3]TCOFs assembled from aza-[3]triangulene core and octa-1,3,5,7-tetrayne linkers. **f,** DFT-PBE band structure for a freestanding $C_8$-A[3]TCOF and corresponding truncated NNN TB Hamiltonian in the Wannier basis. **g,** Molecular model of a $C_2PhC_2$-A[3]TCOFs assembled from aza-[3]triangulene core and 1,4-diethynylbenzene linkers. **h,** DFT-PBE band structure for a freestanding $C_2PhC_2$-A[3]TCOF and corresponding truncated NNN TB Hamiltonian in the Wannier basis.